\documentclass{aa}
\usepackage{graphicx}
\usepackage{natbib}
\usepackage{txfonts}
\bibpunct{(}{)}{;}{a}{}{,} 
\newcommand{\kms}{\rm ~km~s^{-1}}

\newcommand{\cc}{\mbox{ cm}^{-3}}

\newcommand{\ang}{\mbox{ \AA}}
\newcommand{\etal}{\rm et al.~}
\def\EE#1{\times 10^{#1}}

\def\cm3{\rm ~cm^{-3}}
\def\cm2{\rm ~cm^{-2}}
\def\kms{\rm ~km~s^{-1}}

\def\ergsm{\rm ~erg~s^{-1} cm^{-2}}

\def\Ti44{M(^{44}{\rm Ti})}

\def\snr{SNR~0540-69.3}

\def\lsim{\!\!\!\phantom{\le}\smash{\buildrel{}\over
  {\lower2.5dd\hbox{$\buildrel{\lower2dd\hbox{$\displaystyle<$}}\over
                               \sim$}}}\,\,}
\def\gsim{\!\!\!\phantom{\ge}\smash{\buildrel{}\over
  {\lower2.5dd\hbox{$\buildrel{\lower2dd\hbox{$\displaystyle>$}}\over
                               \sim$}}}\,\,}
\def\snr{SN~1987A}

\begin{document}
   \title{High resolution spectroscopy of the line emission from the 
   inner circumstellar ring of SN 1987A and its hot spots\thanks{Based on 
   observations collected at the European Southern Observatory, Chile (ESO programme 70.D-0379).}}


 \author{Per Gr\"oningsson\inst{1}
     	\and
   	Claes Fransson\inst{1}
	\and  
	Peter Lundqvist\inst{1}
	\and  
	Natalia Lundqvist\inst{1}
	\and
        Bruno Leibundgut\inst{2}
	\and 
	Jason Spyromilio\inst{2}
	\and
	Roger A. Chevalier\inst{3}
	\and
	Roberto Gilmozzi\inst{2}
	\and
	Karina Kj\ae r\inst{2}
	\and
	Seppo Mattila\inst{4}
	\and
        Jesper Sollerman\inst{1,5}
    	}

   \offprints{P. Gr\"oningsson}

   \institute{
	Stockholm Observatory, Stockholm University,
              AlbaNova University Center, SE-106 91 Stockholm, Sweden\\
              \email{per@astro.su.se, claes@astro.su.se, peter@astro.su.se}
	\and
	European Southern Observatory,
	Karl-Schwarzschild-Strasse 2, D-85748 Garching, Germany
	\and 
       Department of Astronomy, University of
       Virginia, P.O. Box 400325, Charlottesville, VA 22904, U.S.A.
	\and 
       Astrophysics Research Centre, School of Mathematics and Physics,
			 Queen's University Belfast, BT7 1NN, UK
  \and
       Dark Cosmology Centre, Niels Bohr Institute, University of Copenhagen,
       Juliane Maries Vej 30, 2100 Copenhagen, Denmark
	}

   \date{Received April, 2007; accepted ?}

\titlerunning{Line emission from the inner ring of SN~1987A}
\authorrunning{Gr\"oningsson et al.}

   \abstract{We discuss high resolution VLT/UVES observations (FWHM
   $\sim 6 \kms$) from October 2002 (day $\sim5700$ past explosion) of the shock interaction of SN
   1987A and its circumstellar ring. A nebular analysis of the narrow
   lines from the unshocked gas indicates gas densities of $(\sim1.5-5.0)\EE3 \cc$
   and temperatures of $\sim6.5\EE3-2.4\EE4$ K. This is consistent with the
   thermal widths of the lines. From the shocked component we observe
   a large range of ionization stages from neutral lines
   to [Fe XIV]. From a nebular analysis we find that the density in
   the low ionization region is $4\EE6-10^7 \cc$. There is a clear
   difference in the high velocity extension of the low ionization lines and that of
   lines from [Fe X-XIV], with the latter extending up
   to $\sim -390 \kms$ in the blue wing for [Fe XIV], while the low ionization lines
   extend to typically $\sim -260 \kms$. For H$\alpha$ a faint
   extension up to $\sim -450 \kms$ can be seen probably arising from a small fraction of shocked
   high density clumps. We discuss these observations in the context of
   radiative shock models, which are qualitatively consistent with the
   observations. A fraction of the high ionization lines may originate
   in gas which has yet not had time to cool down, explaining the
   difference in width between the low and high ionization lines. The
   maximum shock velocities seen in the optical lines are $\sim 510 \kms$.
   We expect the maximum width of especially the low ionization lines to increase with time.

   \keywords{supernovae: individual: SN 1987A --
                circumstellar matter, shocks
               }
   }

   \maketitle
	
%

\section{Introduction}
\label{sec:introd}

The core-collapse supernova (SN) 1987A in the Large Magellanic Cloud (LMC) is 
much closer to us than the SNe we normally observe. Detailed 
studies of SN 1987A have therefore contributed tremendously to our 
understanding of this type of SNe. Its interaction with circumstellar gas 
has also contributed to our understanding of shockwave physics \citep[see e.g.][]{rac97,mccray05}.

The radioactive debris of SN 1987A is at the center of a triple ring
system where the two outermost rings are about three times the size of
the innermost ring \citep{WW92}. The
rings are elliptical in shape with the inner ring centered around the
SN and the two outer rings are centered to the north and south of the
SN forming a possible hour-glass structure. Since the inner ring
(henceforth called the equatorial ring, or just ER) appears to be
intrinsically close to circular in shape \citep{GU98,sugerman05}, its observed ellipticity is interpreted as a tilt angle
to the line of sight of $\sim 44^\circ$
\citep{plait95,burrows95,sugerman02}.  How all the rings were formed is
still unknown, but it seems clear that the ER is the result of the
final blue supergiant wind interacting with previously emitted gas in
the form of an asymmetric wind-like structure
\citep[e.g.,][]{BL93,CD95}. In the models by \citet{MP05} the strong asymmetry is explained in a
merger scenario being due to a rotationally enforced outflow. In these models also
the outer rings may be explained by the merger, and the rings are likely
to have been formed during the last $\sim$20,000 years before the explosion. Many details in both the observations and models are, however, uncertain.

The observed ER was photoionized to a high state of ionization by the
SN flash accompanying the SN breakout \citep{FL89}. It has since the
breakout been cooling and recombining, giving rise to a multitude of
narrow emission lines \citep[e.g.,][]{LF96}.  Already early it was
realized that the ER should be reionized $\sim 10-20$ years after the
explosion when the expanding SN ejecta would start to interact with
the ER \citep{luo91,luo94}. This event would cause a rebrightening of
the ring and the first indication of this interaction became visible
as a bright spot (Spot 1) in 1997 on the north side of the ring at PA
= $29^\circ$ \citep{gar97,sonneborn98}.

Modelling of Spot 1 \citep{michael00} suggests that it is an inward
protrusion of the dense \citep[$\sim10^4\cc$,][]{LF96} ER and that it is
embedded in the lower density \citep[$\sim10^2\cc$,][]{CD95,lun99}
H~II region interior to the ring. The shocked gas drives a forward
shock into this H~II region and the bright Spot 1 is the result of the
interaction of the shock and an inward protrusion of the inner
circumstellar ring. When this happened, slower radiative shocks were
transmitted into the protrusion and the radiation from these shocks
appeared as a bright spot.

The velocities of the transmitted shocks are given by $V_s \sim V_b
(n_{\rm HII} / n_{\rm sp})^{\frac{1}{2}}$, where $V_b$ is the blast
wave velocity and $n_{\rm HII}$ and $n_{\rm sp}$ are the densities of
the H~II region and the protrusion, respectively.  The fastest
transmitted shocks are expected for head-on collisions, while slower
velocities occur for tangential shocks. The shock velocities therefore
depend on both the density and geometry of the spot. As a consequence,
the transmitted shocks will have a wide range of velocities ($\sim
10^2-10^3 \kms$). However, only a fraction of these shocks will
contribute to the observed UV/optical emission from the spot \citep{pun02}.

Since 1997 an increasing number of bright spots have been observed in
the ER \citep{sugerman02}.
This has marked the transition between the SN state and the SN remnant
state, and for the first time we have the opportunity to study such a
transition. A central issue for understanding the physics of the
interaction is the hydrodynamics and physical conditions of the
shocked region. High spectral resolution in combination with good
spatial resolution is here invaluable. To address these issues we
present in this paper high resolution optical spectroscopic VLT/UVES
data, taken at 2002 October 4$-$7 (days 5702$-$5705). In a previous paper we have
discussed some special aspects of these data, in particular the
presence of a number of coronal lines from [Fe X-XIV] \citep{gro06}.
Here we discuss the observational analysis in greater detail, and in
particular we concentrate on an analysis of lines from the low and
medium ionization stages, as well as the narrow line component. In a
forthcoming paper we will discuss the time evolution of the line
emission using data from several epochs.

In Sect. 2 we discuss the
observations and data reductions. In Sect. 3 we describe how to
separate out emission lines both from different regions such as
unshocked and shocked gas and different spatial positions of the
ER. The analysis of these data is done in Sect. 4 where we also
describe what conclusions can be drawn from the line profiles and
nebular analysis. The results are discussed and summarized in Sect. 5.


\section{Observations}
\label{sec:obser}

Service mode observations of supernova remnant (SNR) 1987A were obtained on 2002 October
4$-$7 using the cross-dispersed Ultraviolet and Visual Echelle
Spectrograph (UVES) at the ESO Very Large
Telescope\footnote{http://www.eso.org/paranal/} at Paranal, Chile. Details of the
observations are given in Table \ref{tab:obslog}. The spectrograph
separates the light beam into two different arms. One arm covering
the shorter wavelengths ($3030-4990\ang$) has a CCD detector with a
spatial resolution of $0\farcs246$/pix.  The arm covering the longer wavelengths
($4760-10\,600\ang$) has two CCDs with spatial resolutions of
$0\farcs182$/pix. The UVES spectrograph was used with two different
dichroic settings and hence covering in total the wavelength range
between $3030-10\,600\ang$ with gaps at $5770-5830\ang$ and $8540-8650\ang$.

\begin{figure}
\resizebox{\hsize}{!}{\includegraphics{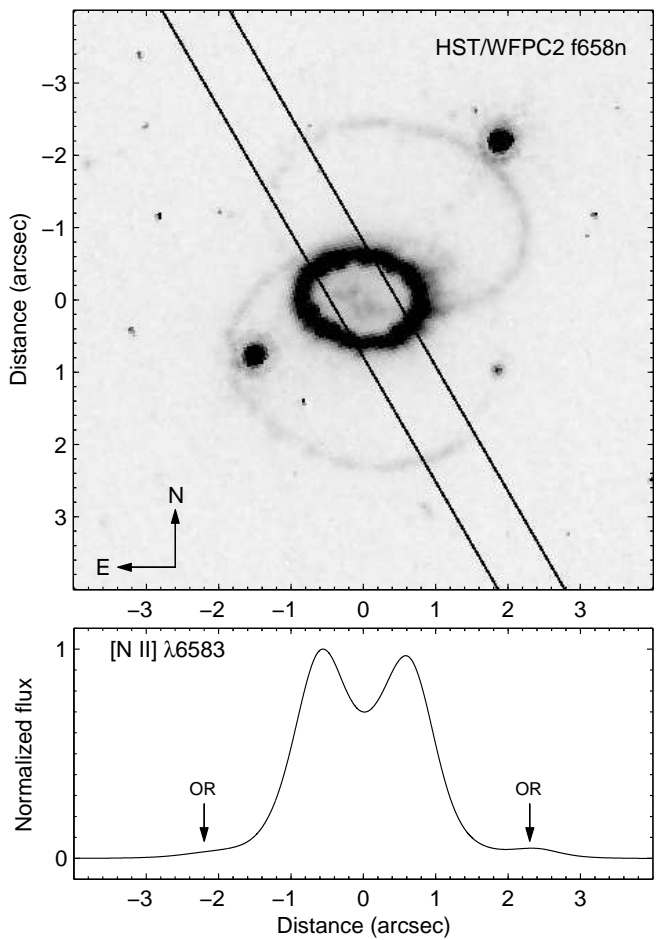}} 
\caption{\textbf{Upper panel:} HST/WFPC2 [N II] $\lambda$6583 image of the 
triple ring system of SN 1987A from 2002 May 10. Overlaid is the slit position 
of our VLT/UVES observations (PA $=30^{\circ}$) from 2002 October. The slit 
width is $0\farcs8$ and the image 
size $8\farcs0\times 8\farcs0$. \textbf{Lower panel:} Flux 
of [N~II] $\lambda$6583 along the slit in our 2002 October spectrum. 
The north (left) and south (right) parts of the inner ring result in two 
prominent peaks. Note also the two outer ring components (marked by ``OR") on 
either side of the inner ring.}
\label{fig:slitpos}
\end{figure}

To encapsulate Spot~1 and to minimize the influence from the bright
nearby stars a $0\farcs8$ wide slit was centered on the SN and rotated
to PA = $30^\circ$. For these reasons and to be
able to separate the north and south parts of the ER we requested an
atmospheric seeing better than $0\farcs8$ for the observations (see
Fig. \ref{fig:slitpos}). Since there are no real point sources inside the slit it is difficult to get a direct measure of the image quality from the data. Instead we chose to use the DIMM as an estimate. It typically overestimates the actual seeing by $\sim0\farcs15$ but can serve as an upper limit. From Table \ref{tab:obslog} we see that the
average DIMM seeing for these observations was $\sim0\farcs8$.
After a careful check we concluded that the data taken also at somewhat
poorer DIMM seeing were good enough for our purposes, which are mainly
to separate the north and south parts of the ER. (See also Appendix 1
for a discussion on the influence of seeing.)

For the slit width used for these observations the resolving power is
$\sim 50\,000$, which corresponds to a spectral resolution of $\sim 6
\kms$.

\subsection{Data reduction}
The calibration data, according to the
UVES calibration plan, include bias, wavelength calibration
spectra, flat fields and spectrophotometric standard stars. Using the UVES pipeline
version 2.0 as implemented in MIDAS, we produced the calibration sets needed
for bias subtraction, flat-fielding, order extraction and wavelength
calibration of the science data.

In order to produce reduced wavelength calibrated 2D spectra, the UVES pipeline 
command REDUCE/SPAT was used. The orders of the reduced and calibrated 2D 
spectra suffer from high noise levels near their ends, and especially so at 
the blue end. Moreover, by comparing consecutive orders of standard stars, we found that the flux levels at the overlapping regions
differed significantly from each other. That could of course affect 
the relative fluxes of lines located near the ends of the orders. Since the automatic merging of the orders within REDUCE/SPAT did not give satisfactory results we chose to do the merging outside the pipeline. We first excluded both ends of each order and then applied a linear weighted 
averaging between the remaining overlapping regions for every order of the 2D frames before
merging them. The overlap after cutting the ends of the orders were $\sim10\ang$ except for reddest part of the spectrum where the overlap was too small for making adequate mergings. Hence, those regions were removed from the final 1D spectrum (see Fig. \ref{fig:fullspec}) The accuracy in the wavelength calibration for the merged 2D 
spectra was checked against strong night sky emission lines and the systematic error was 
found to be $\sim 0.01\ang$ over the whole spectral range, which is also guaranteed by the UVES 
team\footnote{http://www.eso.org/instruments/uves/}.
In order to transform the wavelength calibrated data to a heliocentric frame 
we used the task RVCORRECT within the IRAF package to calculate the 
radial velocity component of the instrument with respect to SN 1987A due to 
the motion of the earth.

In addition to the sky emission lines the resulting 2D spectra show strong
background emission from the LMC. In order to remove this source we
used the IRAF routine BACKGROUND. We concluded that a simple linear interpolation
could not sufficiently reduce the LMC background. Instead we chose a
second order Legendre polynomial interpolation as an estimate of the
background level. This interpolation proved to be a good estimate for the background emission in
most cases. However, for [O~II] and [O~III] where the background is
strong, the procedure was not able to remove all the background
emission and some background artifacts in form of weak emission features may still remain after the
subtraction.  Nevertheless, since the background emission component
has a different systemic velocity than the SN most of it is spectrally
resolved from the ring emission, and does not significantly affect our
results.

To extract 1D spectra from the 2D spectra we fitted the double peaked
spatial profile (see Fig. \ref{fig:slitpos}) with the sum of two Gaussians. For
the best fit, i.e. the least-square solution, each Gaussian represents one part of the ER i.e. north or
south. The two-gaussian model was then extracted in the dispersion
direction. The resulting 1D spectra were average-combined with weights
proportional to their exposure times. Within this process we applied a sigma clipping routine which effectively rejected outliers of the data such as cosmic rays which heavily contaminated the individual spectra.

The flux calibration of the data was performed by using the UVES
master response curves and the atmospheric extinction curve provided
by ESO.  Different response curves for different standard settings and
epochs are provided. To check the quality of the flux calibration we
reduced the spectrophotometric standard star spectra with the same calibration sets
as for the science spectra. The width of the slit used for the standard star observations was $10\farcs0$ and slit losses should therefore be negligible. The extracted flux-calibrated
standard-star spectra was compared to their tabulated physical
fluxes. The fluxes obtained by using the master response curves showed
to be accurate to $\sim 10\%$ in level and shape in general. The science spectra were corrected for this discrepancy.

The data were observed with an atmospheric
dispersion corrector and thus we consider the slit losses to be close
to monochromatic. However, another source of
uncertainties could be due to possible misalignment of the relatively narrow
slit ($0\farcs8$) for the science observations which could cause
substantial losses in flux. Investigations by eye of the slit images,
provided as a part of the UVES calibration plan, revealed that the slit was
accurately placed across the SN.  Nevertheless, flux losses are still,
of course, seeing dependent. To estimate how sensitive fluxes are to different
seeing conditions we compared various line fluxes of the individual
science exposures and concluded that the fluxes differ by $\lsim 10\%$
between the different exposures. From these measurements, and the
results in Appendix 1 where we have compared our UVES fluxes to data
taken with the Hubble Space Telescope at similar epochs, we are
confident that the systematic error of the absolute flux should be less than $20-30$\%. As discussed above, the {\it relative} fluxes should, however, be accurate to $\sim 10-15$\%.

To create the final spectrum  the
combined spectra from all settings were merged by applying a linearly
weighted averaging between the overlapping regions for the settings.
\begin{table*}
\centering
\caption{VLT/UVES observations of SN 1987A and its rings at days 5702$-$5705 past explosion. The airmasses of these observations were $\sim1.5$.}
\begin{tabular}{c c c c c c}
\hline
\hline
Date & Grating Setting & Wavelength Range & Slit & Seeing  &Exposure time \\ & & (nm) & (arcsec) & (arcsec) & (s) \\
\hline

2002-10-04 & CD\#4 860 & 660 - 1060 & $12.0\times 0.8$ &$0.91\pm0.06$&3600 \\
2002-10-04 & CD\#2 437 & 373 - 499 & $10.0\times 0.8$ &$0.91\pm0.06$&3600 \\
2002-10-04 & CD\#2 437 & 373 - 499 & $10.0\times 0.8$ &$1.05\pm0.20$&2160 \\
2002-10-04 & CD\#4 860 & 660 - 1060 & $12.0\times 0.8$ &$1.05\pm0.20$&2160 \\
2002-10-05 & CD\#4 860 & 660 - 1060 & $12.0\times 0.8$ &$0.44\pm0.05$&3600 \\
2002-10-05 & CD\#2 437 & 373 - 499 & $10.0\times 0.8$ &$0.44\pm0.05$&3600 \\
2002-10-06 & CD\#3 580 & 476 - 684 & $12.0\times 0.8$ &$0.99\pm0.13$&3600 \\
2002-10-06 & CD\#1 346 & 303 - 388 & $10.0\times 0.8$ &$0.99\pm0.13$&3600 \\
2002-10-06 & CD\#3 580 & 476 - 684 & $12.0\times 0.8$ &$0.94\pm0.18$&3600 \\
2002-10-06 & CD\#1 346 & 303 - 388 & $10.0\times 0.8$ &$0.94\pm0.18$&3600 \\
2002-10-07 & CD\#1 346 & 303 - 388 & $10.0\times 0.8$ &$0.65\pm0.05$&3000 \\
2002-10-07 & CD\#3 580 & 476 - 684 & $12.0\times 0.8$ &$0.65\pm0.05$&3000 \\
\hline
\label{tab:obslog}
\end{tabular}
\end{table*}

\begin{figure*}
\resizebox{\hsize}{!}{\includegraphics{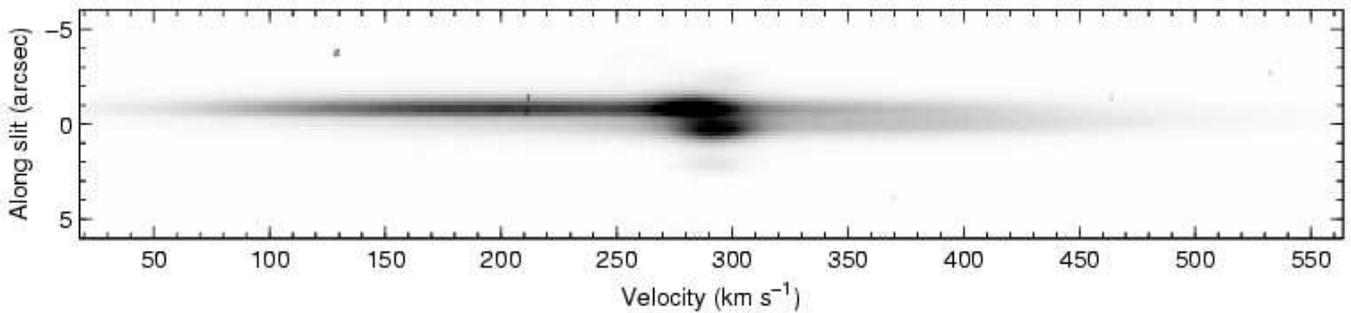}} 
\caption{The 2D background subtracted spectrum from 2002 October 6 showing
H$\alpha$. The vertical direction of the plot is along the slit, with
the northern part in the upward direction. The total spatial extent is
12 arcsec. The velocities are with respect to the rest wavelength of H$\alpha$. Note how the two intersections of the slit with the ER and the outer rings are clearly
visible around the rest velocity of the supernova, as well as the weaker
emission from the outer rings. The broader (few$\times 10^2 \kms$)
emission (the ``intermediate" component) is asymmetric, with the blue
side dominating for the northern ER, and the red side for the southern
ER.}
\label{fig:twod}
\end{figure*}


\section{Results}
\begin{figure*}
\resizebox{\hsize}{!}{\includegraphics{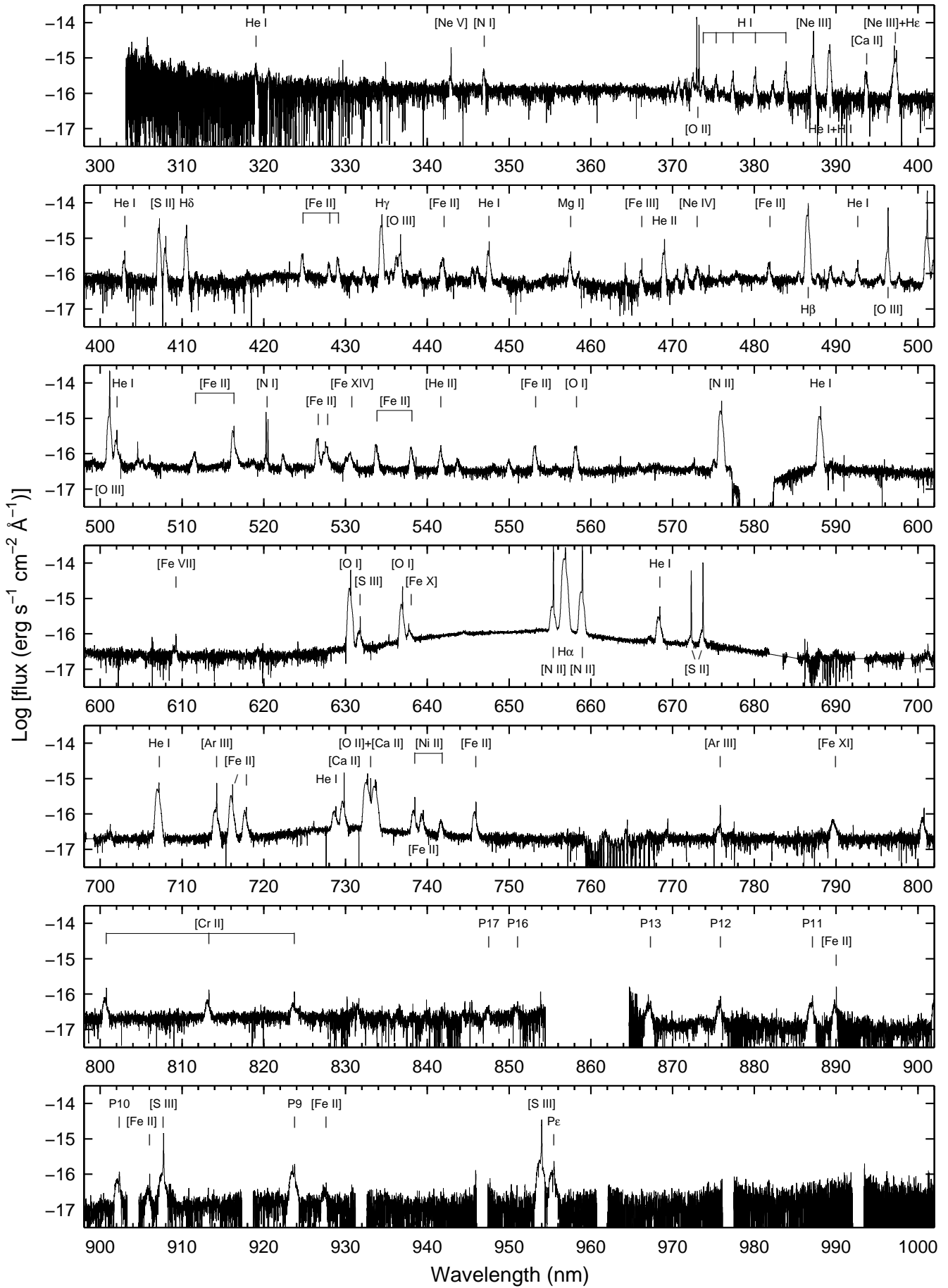}}
\caption{Extracted 1D spectrum from the northern part of the ER. The
most significant/important lines are marked. The gaps in the reddest
part of the spectrum (lowest panel) are due to the poor overlapping in
wavelength between different orders. Many of the lines show both
narrow and intermediate components (cf. Tables
\ref{tab:narrowlines}--\ref{tab:broadlinesS}). Note also the very
broad H$\alpha$ emission component (the middle panel) originating from the
ejecta (in the following called the ``broad" component).}
\label{fig:fullspec}
\end{figure*}

From our data we were able to spatially
separate the northern and the southern parts of the ER in the reduced
2D frames (see Fig. \ref{fig:twod}) to extract 1D spectra.
Figure \ref{fig:fullspec} shows the full spectrum from the northern
part of the ring, with a number of the stronger lines marked.

First we note the very broad H$\alpha$ line, with an extension of
$\sim15\,000\kms$. As discussed by \citet{smith05,heng06}, this most
likely originates at the reverse shock, propagating back into the
ejecta. Also, at $\sim7300\ang$ there is a broad component most likely
from [Ca~II]$\lambda\lambda$7292,7324. These will be discussed in
detail in a future paper.

\begin{figure}
\resizebox{\hsize}{!}{\includegraphics{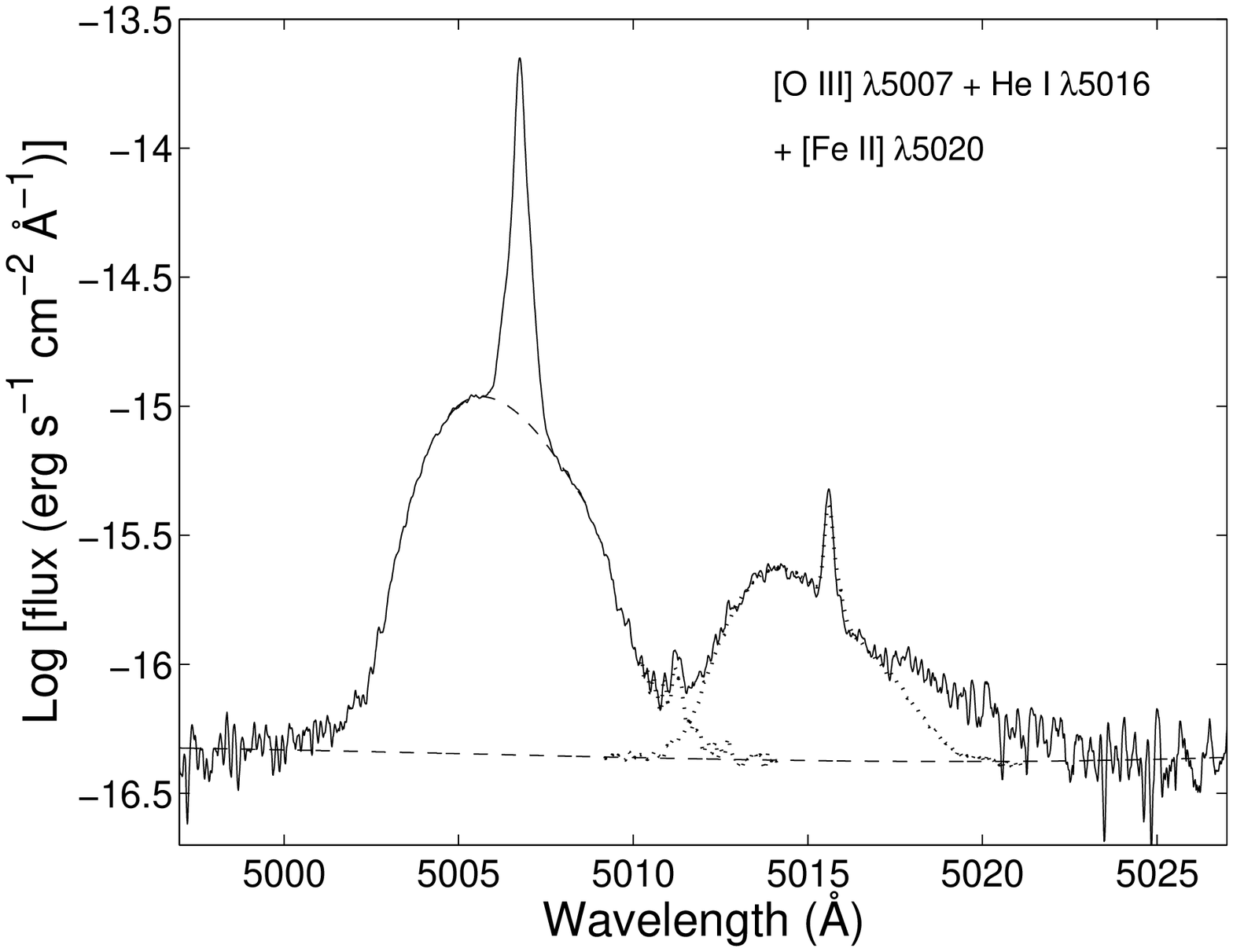}} 
\caption{The intermediate and narrow components of [O
III]~$\lambda$5006.8 and He I~$\lambda$5016 for the northern part of the ER.  The dashed
lines show the estimated zero intensity levels for the narrow and
intermediate components, respectively. The red wing of the intermediate
component of [O~III] is blended with He I~$\lambda$ 5015.7. He I~$\lambda$ 5875.6 is used as a template for He I~$\lambda$ 5015.7 and the dotted line shows the result of the deblending procedure. Note also [Fe II]~$\lambda$5020.2 that gives the excess flux in the red wing of He I~$\lambda$ 5015.7. The wavelength scale is corrected for redshift.}
\label{fig:OIII_profile}
\end{figure}

Most of the emission lines show a narrow component (FWHM $\sim
10\kms$) on top of a broader (HWZI $\sim 300\kms$, henceforth called
the ``intermediate" component) (see Fig. \ref{fig:OIII_profile}). We
interpret the narrow component as the emission from the unshocked CSM
and the intermediate component to come from the shocked CSM \citep[see
also][]{pun02}. To separate the emission from these two components we
have made an interpolation of the intermediate component using least
squares spline approximations (see Fig. \ref{fig:OIII_profile}).
Another problem that had to be addressed was the blending of different intermediate line components. In Fig. \ref{fig:OIII_profile} we show a typical example of how the deblending process was done. Here [O~III] $\lambda$5007 was separated from He I $\lambda$5016 by fitting and scaling He I $\lambda$5876 to the latter line. Assuming that these two He I lines have similar profiles, the He I $\lambda$5016 line could then easily be subtracted from [O~III] $\lambda$5007.

\begin{figure}
\resizebox{\hsize}{!}{\includegraphics{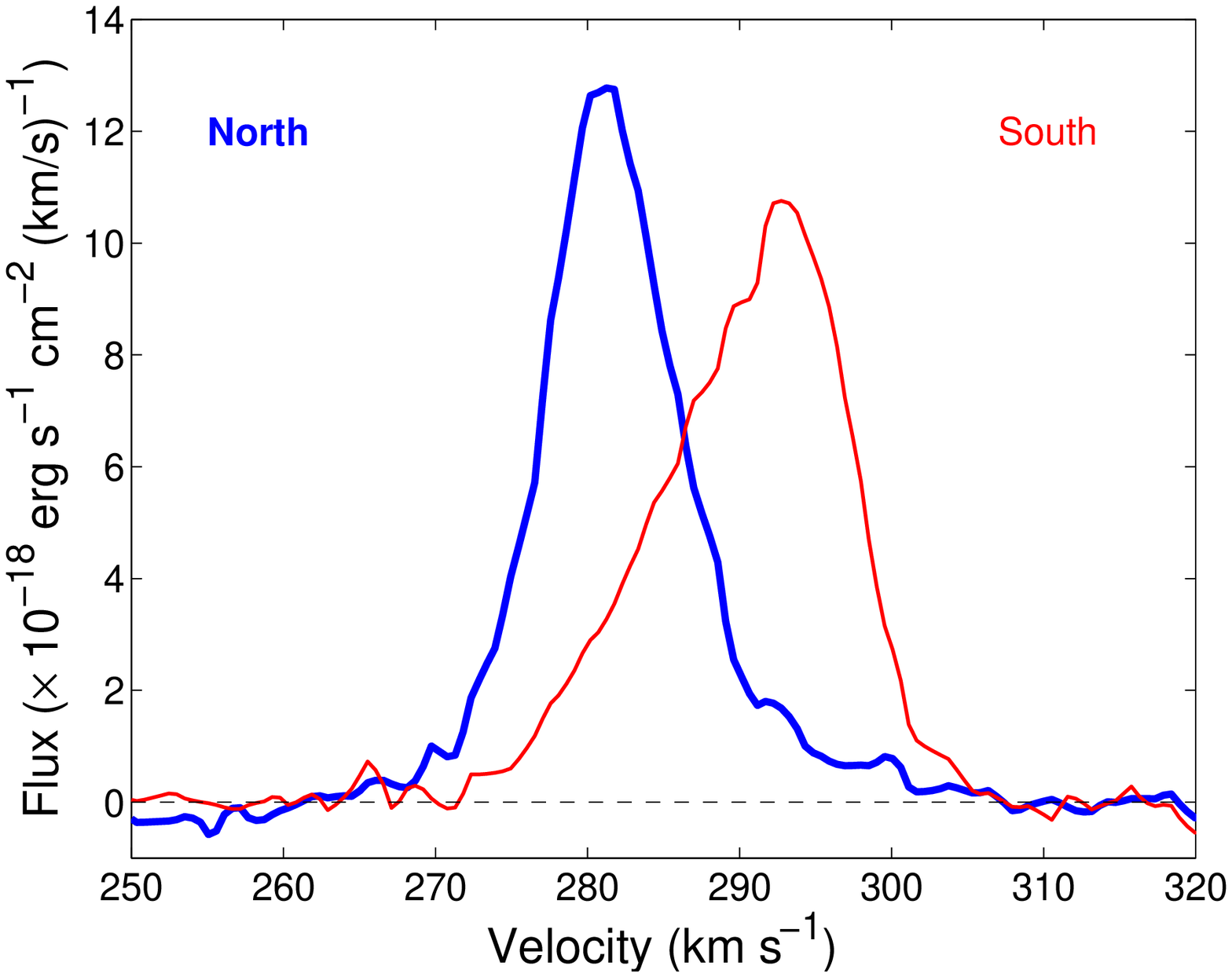}}
\caption{[Fe~II] $\lambda$7155 line profiles from the northern (thick line) and
southern (thin line) parts of the ER. The velocity 
scale is heliocentric. The lines appear as skewed Gaussians due to contributions 
from the opposite side of the ring in both cases. See text for more details.}
\label{fig:FeII_helio}
\end{figure}

The line profile of the narrow components is mainly due to thermal
broadening at least for the lightest elements such as H and
He. However, since the ER is an extended object, the observed profile
will deviate from a simple Gaussian, looking more like a skewed
Gaussian (see Fig. \ref{fig:FeII_helio}). By contrast, the line profile of the intermediate component
is dominated by shock dynamics and there is no reason for it to be
Gaussian at all. Figure \ref{fig:OIII_profile} shows [O III]~$\lambda$5007 and He I~$5016$ for the two
components. In Sect. 4 we will discuss the physical cause of the line
profiles in more detail.

Identifications of all emission lines are given in Table
\ref{tab:ions}, and measurements of the velocities and fluxes of the most significant emission
lines are presented in
Tables \ref{tab:narrowlines}--\ref{tab:broadlinesS}. The values listed
in the tables were obtained by fitting a sum of Gaussians to the line
profiles and the background was estimated by a third-order polynomial interpolation. The values given in brackets are the statistical $1\sigma$
uncertainties of these fits. We note, however, that the systematic errors in the relative fluxes are $10-15$\%, which in many cases dominate over the statistical errors.

The velocity ranges of the intermediate components ($V_{\rm blue}$ and $V_{\rm red}$)
are defined as the ranges where the flux is more than $5 \%$ of the
peak flux. The reason for using this, rather than the maximum extent
of the line, is that the latter is very sensitive to the signal-to-noise ratio (S/N) of the
line, which varies by a factor of $\ga 100$. With this measure we can
therefore better compare lines of very different fluxes. 
The maximum extent is discussed in
Sect. \ref{sec:lprof}.

The peak velocities for the narrow components are with respect to the
heliocentric motion, whereas {\it the peak velocities for the
intermediate components are relative to the peak velocity of the
corresponding narrow component}. This provides a natural frame for the
ejecta impact on the ER. For the extinction we adopted a reddening of
$E(B-V)=0.16{\rm ~mag}$ \citep{FW90} with $E(B-V)=0.06{\rm ~mag}$ from the Milky way
\citep[e.g.,][]{SS03} and $E(B-V)=0.10{\rm ~mag}$ from the LMC. The reddening
law was taken from \cite{Card89} using $R_{V}=3.1$.

\begin{table}
\centering
\caption{The dereddened Balmer line fluxes relative to H$\beta$ for the unshocked gas. Only the 1$\sigma$ statistical errors are given. Systematic errors are discussed in the text.}
\begin{tabular}{c c c c c}
\hline
\hline
Position & H$\alpha$ & H$\beta$  & H$\gamma$ & H$\delta$ \\
\hline
north & $354.2\pm1.1$ & $100$ & $47.0\pm0.3$ & $22.5\pm0.2$\\
south & $322.8\pm1.0$ & $100$ & $43.5\pm0.3$ & $17.9\pm0.1$\\
Case B& $287$ & $100$ & $46.6$ & $25.6$\\
\hline
\label{tab:balmerflux}
\end{tabular}
\end{table}

\begin{table*}
\centering
\caption{Emission lines grouped by ions.}
\begin{tabular}{l l}
\hline
\hline
Ion & Wavelengths ($\ang$)\\
\hline
H I & 3691.6b, 3697.2, 3703.9, 3712.0, 3721.9, 3734.4,  3750.2, 3770.6, 3797.9, 3835.4, 3889.0, 3970.1, 4101.7, 4340.5,\\
\smallskip
&4861.3, 6562.8, 8438.0, 8467.3, 8502.5, 8665.0, 8750.5, 8862.8, 9014.9, 9229.0, 9546.0, 10049.4\\
He I & 3187.7, 3705.0, 3819.6, 3888.7, 4026.2, 4120.8b, 4143.8, 4387.9, 4471.5, 4713.2, 4921.9, 5015.7, 5047.7, 5875.7,\\
\smallskip
&6678.2, 6855.9b, 7065.7, 7281.4, 8444.5a, 8530.9a, 8662.2a\\
\smallskip
He II & 3203.1b, 3710.4a, 4541.6b, 4685.7, 5411.5, 5431.8b, 6527.1a, 8236.8a\\
\smallskip
C I & 8727.1b\\
\smallskip
N I & 3466.5b, 5197.9, 5200.3\\
\smallskip
N II & 5754.5, 6547.9, 6583.3\\
\smallskip
O I & 5577.3, 6300.3, 6363.8, 7254.2-5a, 8446.3-8a\\
\smallskip
O II & 3726.0a, 3728.8a, 7318.9, 7320.0, 7329.7,7330.7\\
\smallskip
O III & 4363.2, 4958.9, 5006.8\\
\smallskip
Ne III & 3868.8, 3967.5\\
\smallskip
Ne IV & 4714.2, 4725.7b\\
\smallskip
Ne V & 3345.9a, 3425.9\\
\smallskip
Mg I & 4562.7a, 4571.1\\
\smallskip
Si II & 5041.0a, 5056.0a, 6347.1a, 6371.4a\\
\smallskip
S II & 4068.6, 4076.4, 6716.4, 6730.8\\
\smallskip
S III & 6312.1, 9068.6, 9530.6\\
\smallskip
Cl II & 9123.6a\\
\smallskip
Ar III & 7135.8, 7751.1\\
\smallskip
Ar IV & 4711.3a, 4740.1a\\
\smallskip
Ar V & 7005.7a\\
\smallskip
Ca II & 3933.7, 7291.5, 7323.9\\
\smallskip
Ca V & 5309.1a, 6086.4a\\
\smallskip
Cr II & 4580.8, 4581.1, 8000.1, 8125.3, 8229.7, 8308.5, 8357.6\\
Fe II & 4114.5b, 4177.2b, 4244.0, 4276.8, 4287.4, 4305.9b, 4319.6b, 4346.9b, 4352.8b, 4358.4b, 4359.3a, 4372.4b, 4416.3,\\
&4452.1b, 4457.9b, 4639.7b, 4774.7b, 4798.3b, 4814.5, 4874.5b, 4889.6, 4905.3, 4950.7b, 4973.4b, 5072.4b, 5111.6,\\
&5158.8, 5220.1, 5261.6, 5268.9b, 5273.3, 5296.8b, 5333.6, 5376.5, 5433.1b, 5477.2b, 5494.3b, 5527.6, 5721.3, 5745.7,\\
\smallskip
&6440.4b, 6872.2b, 6896.2b, 6944.9b, 7155.2, 7172.0, 7388.2, 7452.5, 7637.5, 7686.9, 8891.9, 9051.9, 9226.6, 9267.6\\
\smallskip
Fe III & 4658.1, 4701.5, 4754.7, 4769.4b, 4881.0a, 5011.3a, 5270.4a\\
\smallskip
Fe V & 4227.2a\\
\smallskip
Fe VI & 4972.5a, 5277.8, 5335.2a\\
\smallskip
Fe VII & 3586.3b, 4893.4a, 5720.7, 6087.0\\
\smallskip
Fe X & 6374.5b\\
\smallskip
Fe XI & 7891.8b\\
\smallskip
Fe XIV & 5302.9b\\
\smallskip
Ni II & 6666.8b, 7377.8, 7411.6b\\
\hline
\label{tab:ions}
\end{tabular}
\begin{list}{}{}
 	\item[a] Only a narrow component is detected.
	\item[b] Only an intermediate component is detected.
\end{list}
\end{table*}

\begin{table*}
\centering
\caption{Emission lines from the unshocked gas of the ER.
}
\begin{tabular}{lclcclccc}
\hline
\hline
&&&North&&&South&&Extinction\\
Emission & Rest wavel.& Relative flux$^{\mathrm{a}}$ & $V_{\rm peak}^{\mathrm{b}}$&$V_{\rm fwhm}$ & Relative flux$^{\mathrm{a}}$ & $V_{\rm peak}^{\mathrm{b}}$&$V_{\rm fwhm}$& correction$^{\mathrm{c}}$\\
 &$(\ang)$& &($\kms$)&$(\kms$)& &($\kms$)&$(\kms)$& \\
\hline
\lbrack Ne V\rbrack&3425.86&$\phantom{00}17.2\pm\phantom01.7$&$281.7\pm1.2$&$23.45\pm0.58$&$\phantom{00}13.8\pm\phantom01.6$&$294.4\pm1.0$&$19.55\pm0.57$&2.06\\
\lbrack O II\rbrack&3726.03&$\phantom0109.2\pm\phantom04.1$&$283.0\pm0.4$&$13.82\pm0.13$&$\phantom0101.1\pm\phantom04.0$&$293.3\pm0.5$&$14.77\pm0.23$&1.98\\
\lbrack O II\rbrack&3728.82&$\phantom{00}66.4\pm\phantom03.7$&$280.1\pm0.5$&$14.19\pm0.20$&$\phantom{00}63.8\pm\phantom03.9$&$290.6\pm0.6$&$14.87\pm0.37$&1.98\\
\lbrack Ne III\rbrack&3868.75&$\phantom{00}44.0\pm\phantom01.8$&$281.3\pm0.5$&$15.72\pm0.18$&$\phantom{00}24.7\pm\phantom01.8$&$292.1\pm0.7$&$17.64\pm0.28$&1.94\\
\lbrack S II\rbrack&4068.60&$\phantom{00}17.0\pm\phantom00.9$&$281.3\pm0.4$&$11.07\pm0.15$&$\phantom{00}15.2\pm\phantom00.8$&$291.3\pm0.6$&$13.77\pm0.22$&1.90\\
\lbrack S II\rbrack&4076.35&$\phantom{000}5.4\pm\phantom00.6$&$281.7\pm0.6$&$11.46\pm0.30$&$\phantom{000}5.1\pm\phantom00.6$&$292.2\pm0.9$&$13.60\pm0.44$&1.90\\
H$\delta$&4101.73&$\phantom{00}20.3\pm\phantom00.8$&$281.7\pm1.0$&$28.43\pm0.36$&$\phantom{00}15.6\pm\phantom00.6$&$290.9\pm0.8$&$24.72\pm0.29$&1.89\\
H$\gamma$&4340.46&$\phantom{00}44.6\pm\phantom01.7$&$281.1\pm1.0$&$25.86\pm0.26$&$\phantom{00}39.4\pm\phantom01.6$&$291.0\pm0.8$&$26.52\pm0.24$&1.84\\
\lbrack O III\rbrack&4363.21&$\phantom{00}10.6\pm\phantom00.7$&$279.3\pm0.7$&$17.55\pm0.31$&$\phantom{000}6.7\pm\phantom00.6$&$290.8\pm1.0$&$23.97\pm0.64$&1.84\\
He II &4685.7\phantom0&$\phantom{00}10.1\pm\phantom01.4$&$283.0\pm1.2$&$21.56\pm1.27$&$\phantom{000}7.1\pm\phantom00.7$&$295.6\pm1.6$&$30.39\pm1.41$&1.75\\
H$\beta$&4861.32&$\phantom0100.0$&$280.9\pm0.5$&$26.04\pm0.11$&$\phantom{00}97.9\pm\phantom02.6$&$290.6\pm0.6$&$27.36\pm0.16$&1.71\\
\lbrack O III\rbrack&4958.91&$\phantom{00}81.3\pm\phantom02.1$&$281.0\pm0.4$&$15.69\pm0.11$&$\phantom{00}50.1\pm\phantom01.3$&$291.8\pm0.5$&$22.38\pm0.14$&1.68\\
\lbrack O III\rbrack&5006.84&$\phantom0247.2\pm\phantom05.1$&$281.1\pm0.3$&$15.88\pm0.06$&$\phantom0154.9\pm\phantom03.5$&$292.5\pm0.5$&$22.85\pm0.10$&1.67\\
\lbrack O I\rbrack&5577.34&$\phantom{000}0.5\pm\phantom00.2$&$283.4\pm1.4$&$14.66\pm1.17$&$\phantom{000}0.5\pm\phantom00.2$&$289.5\pm1.4$&$15.19\pm1.23$&1.57\\
\lbrack N 
II\rbrack&5754.59&$\phantom{00}20.6\pm\phantom00.6$&$281.5\pm0.3$&$13.69\pm0.09$&$\phantom{00}19.3\pm\phantom00.5$&$292.0\pm0.4$&$16.21\pm0.16$&1.54\\
He I&5875.63&$\phantom{00}17.3\pm\phantom00.6$&$280.8\pm0.5$&$16.85\pm0.15$&$\phantom{00}16.9\pm\phantom00.6$&$290.3\pm0.5$&$18.63\pm0.16$&1.53\\
\lbrack O I\rbrack&6300.30&$\phantom{00}48.0\pm\phantom01.9$&$281.4\pm0.4$&$11.67\pm0.12$&$\phantom{00}47.7\pm\phantom01.4$&$291.2\pm0.4$&$14.12\pm0.16$&1.47\\
\lbrack S III\rbrack&6312.06&$\phantom{000}2.6\pm\phantom00.3$&$280.6\pm0.7$&$12.78\pm0.47$&$\phantom{000}1.3\pm\phantom00.2$&$291.9\pm1.3$&$18.49\pm0.85$&1.47\\
\lbrack O I\rbrack&6363.78&$\phantom{00}16.1\pm\phantom00.6$&$281.1\pm0.3$&$11.40\pm0.09$&$\phantom{00}16.7\pm\phantom00.4$&$291.6\pm0.4$&$14.50\pm0.12$&1.47\\
\lbrack N II\rbrack&6548.05&$\phantom0359.9\pm\phantom07.3$&$281.9\pm0.3$&$12.95\pm0.05$&$\phantom0352.7\pm\phantom06.9$&$292.5\pm0.3$&$14.88\pm0.05$&1.45\\
H$\alpha$&6562.80&$\phantom0419.2\pm\phantom06.9$&$280.9\pm0.4$&$28.69\pm0.07$&$\phantom0369.8\pm\phantom05.7$&$290.5\pm0.3$&$27.82\pm0.05$&1.45\\
\lbrack N II\rbrack&6583.45&$1112.7\pm18.7$&$280.7\pm0.2$&$12.74\pm0.03$&$1096.4\pm18.2$&$291.2\pm0.2$&$14.99\pm0.04$&1.45\\
He I &6678.15&$\phantom{000}5.5\pm\phantom00.6$&$281.1\pm0.7$&$16.07\pm0.35$&$\phantom{000}4.7\pm\phantom00.3$&$291.6\pm0.7$&$18.12\pm0.32$&1.44\\
\lbrack S II\rbrack&6716.44&$\phantom{00}62.2\pm\phantom01.5$&$281.9\pm0.3$&$11.61\pm0.06$&$\phantom{00}69.6\pm\phantom01.8$&$292.9\pm0.3$&$13.75\pm0.07$&1.43\\
\lbrack S II\rbrack&6730.82&$\phantom0103.5\pm\phantom02.4$&$281.7\pm0.3$&$11.46\pm0.06$&$\phantom0115.5\pm\phantom03.3$&$292.7\pm0.3$&$13.82\pm0.08$&1.43\\
\lbrack Ar III\rbrack&7135.79&$\phantom{000}8.3\pm\phantom00.4$&$279.9\pm0.4$&$12.91\pm0.15$&$\phantom{000}5.4\pm\phantom00.4$&$290.1\pm0.7$&$18.11\pm0.29$&1.39\\
\lbrack Fe II\rbrack&7155.16&$\phantom{000}4.8\pm\phantom00.3$&$281.3\pm0.4$&$\phantom09.91\pm0.13$&$\phantom{000}4.9\pm\phantom00.2$&$293.2\pm0.4$&$13.73\pm0.28$&1.39\\
\lbrack O II\rbrack&7319$^{\mathrm{d}}$\phantom{00}&$\phantom{00}13.1\pm\phantom01.0$&-&-&$\phantom{00}12.3\pm\phantom00.7$&-&-&1.37\\
\lbrack Ca II\rbrack&7323.89&$\phantom{000}7.2\pm\phantom00.3$&$279.9\pm0.4$&$\phantom09.77\pm0.13$&$\phantom{000}9.8\pm\phantom00.3$&$291.9\pm0.4$&$13.39\pm0.21$&1.37\\
\lbrack O II\rbrack&7330$^{\mathrm{e}}\phantom{00}$&$\phantom{00}11.4\pm\phantom00.8$&-&-&$\phantom{000}9.5\pm\phantom00.8$&-&-&1.37\\
\lbrack S III\rbrack&9068.6\phantom0&$\phantom{00}21.5\pm\phantom00.7$&-&$12.76\pm0.10$&$\phantom{00}11.2\pm\phantom00.4$&-&$18.90\pm0.17$&1.24\\
\lbrack S III\rbrack&9531.10&$\phantom{00}55.2\pm\phantom02.0$&-&$12.84\pm0.12$&$\phantom{00}27.8\pm\phantom00.9$&-&$17.58\pm0.22$&1.22\\
\hline
\label{tab:narrowlines}
\end{tabular}
\begin{list}{}{}
 \item[$^{\mathrm{a}}$] Fluxes are relative to the flux of H$\beta$ in
  the Northern ER: ``100" corresponds
 to $(31.3\pm 0.5) \times 10^{-16}\ergsm$. Fluxes are {\bf not}
  corrected for the extinction. The errors are only statistical, and
  do not include systematic uncertaities (see text).
 \item[$^{\mathrm{b}}$] The recession velocity of the peak flux.
 \item[$^{\mathrm{c}}$] $E(B-V)=0.16{\rm ~mag}$, with $E(B-V)=0.10{\rm ~mag}$ for LMC and $E(B-V)=0.06{\rm ~mag}$ for the Milky Way.
 \item[$^{\mathrm{d}}$] Applies to both lines at 7318.92 \AA\ and 7319.99 \AA.
 \item[$^{\mathrm{e}}$] Applies to both lines at 7329.67 \AA\ and 7330.73 \AA.
 \end{list}
\end{table*}

\begin{table*}
\centering
\caption{Emission lines from the shocked gas of the northern ER.
}
\begin{tabular}{lclcccc}
\hline
\hline
Emission &Rest wavel.& Relative flux$^{\mathrm{a}}$ &$V_{\rm peak}^{\mathrm{b}}$&$V_{\rm blue}^{\mathrm{c}}$&$V_{\rm red}^{\mathrm{c}}$&Extinction\\
 &$(\ang)$& &($\kms$)&($\kms$)&($\kms$)&correction$^{\mathrm{d}}$\\
\hline
\lbrack Ne V\rbrack&3425.86&$\phantom{00}14.8\pm\phantom03.6$&$-136.3\pm21.4$&$-259.1\pm34.9$&$202.9\pm43.3$&2.06\\
\lbrack Ne III\rbrack&3868.75&$\phantom0111.1\pm\phantom03.6$&$-\phantom087.8\pm\phantom06.2$&$-263.1\pm\phantom04.0$&$216.9\pm\phantom08.9$&1.94\\
\lbrack S II\rbrack&4068.60&$\phantom0171.5\pm\phantom04.6$&$-\phantom077.2\pm\phantom04.3$&$-240.5\pm\phantom04.7$&$180.4\pm\phantom05.5$&1.90\\
\lbrack S II\rbrack&4076.35&$\phantom{00}45.7\pm\phantom02.1$&$-\phantom076.9\pm\phantom06.3$&$-254.9\pm\phantom05.3$&$203.0\pm11.2$&1.90\\
H$\delta$&4101.73&$\phantom0105.5\pm\phantom02.6$&$-\phantom086.8\pm\phantom04.7$&$-261.4\pm\phantom02.4$&$197.5\pm\phantom06.6$&1.89\\
H$\gamma$&4340.46&$\phantom0223.4\pm\phantom05.0$&$-\phantom088.4\pm\phantom04.2$&$-260.0\pm\phantom02.5$&$180.5\pm\phantom04.1$&1.84\\
\lbrack O III\rbrack&4363.21&$\phantom{00}32.0\pm\phantom02.0$&$-\phantom067.5\pm\phantom07.0$&$-251.5\pm\phantom08.0$&$187.7\pm15.1$&1.84\\
\lbrack Fe III\rbrack&4658.05&$\phantom{000}9.8\pm\phantom01.1$&$-101.2\pm11.5$&$-270.9\pm17.3$&$204.9\pm37.8$&1.76\\
He II&4685.7\phantom0&$\phantom{00}37.3\pm\phantom01.7$&$-\phantom094.7\pm\phantom06.8$&$-265.9\pm\phantom05.8$&$182.8\pm13.1$&1.75\\
H$\beta$&4861.32&$\phantom0524.4\pm\phantom08.2$&$-\phantom085.9\pm\phantom03.2$&$-258.7\pm\phantom01.1$&$189.1\pm\phantom01.8$&1.71\\
\lbrack O III\rbrack&4958.91&$\phantom{00}45.7\pm\phantom01.7$&$-\phantom066.1\pm\phantom07.3$&$-243.7\pm\phantom05.6$&$179.4\pm\phantom06.9$&1.68\\
\lbrack O III\rbrack&5006.84&$\phantom0140.0\pm\phantom02.7$&$-\phantom058.5\pm\phantom04.4$&$-252.5\pm\phantom02.9$&$212.8\pm\phantom03.3$&1.67\\
\lbrack Fe XIV\rbrack&5302.86&$\phantom{00}12.2\pm\phantom01.0$&$-132.8\pm11.5$&$-386.7\pm10.7$&$352.3\pm46.7$&1.61\\
\lbrack O I\rbrack&5577.34&$\phantom{00}16.6\pm\phantom01.0$&$-\phantom090.3\pm\phantom07.0$&$-253.7\pm\phantom05.2$&$194.4\pm16.9$&1.57\\
\lbrack N II\rbrack&5754.59&$\phantom0223.7\pm\phantom03.7$&$-\phantom089.6\pm\phantom03.4$&$-252.7\pm\phantom01.3$&$187.3\pm\phantom02.0$&1.54\\
He I&5875.63&$\phantom0156.1\pm\phantom02.6$&$-\phantom084.3\pm\phantom03.5$&$-253.3\pm\phantom01.4$&$191.5\pm\phantom02.3$&1.53\\
\lbrack Fe VII\rbrack&6087.0\phantom0&$\phantom{000}2.3\pm\phantom00.4$&$-102.6\pm16.4$&$-345.6\pm22.4$&$139.9\pm22.2$&1.50\\
\lbrack O I\rbrack&6300.30&$\phantom0237.4\pm\phantom03.7$&$-\phantom076.9\pm\phantom02.0$&$-229.9\pm\phantom00.8$&$169.0\pm\phantom01.8$&1.47\\
\lbrack S III\rbrack&6312.06&$\phantom{00}11.4\pm\phantom00.8$&$-\phantom092.8\pm\phantom09.3$&$-254.1\pm\phantom09.1$&$185.2\pm18.9$&1.47\\
\lbrack O I\rbrack&6363.78&$\phantom{00}81.3\pm\phantom02.6$&$-\phantom076.6\pm\phantom04.2$&$-233.0\pm\phantom05.2$&$184.0\pm\phantom07.2$&1.47\\
\lbrack Fe X\rbrack&6374.51&$\phantom{000}9.8\pm\phantom00.8$&$-127.8\pm10.9$&$-348.7\pm11.7$&$156.2\pm20.7$&1.47\\
\lbrack N II\rbrack&6548.05&$\phantom{00}75.6\pm\phantom01.7$&$-\phantom080.8\pm\phantom04.5$&$-247.4\pm\phantom02.7$&$178.6\pm\phantom03.5$&1.45\\
H$\alpha$&6562.80&$2112.2\pm31.0$&$-\phantom082.1\pm\phantom02.7$&$-259.8\pm\phantom00.9$&$195.4\pm\phantom00.9$&1.45\\
\lbrack N II\rbrack&6583.45&$\phantom0227.5\pm\phantom03.7$&$-\phantom078.2\pm\phantom03.2$&$-244.0\pm\phantom01.5$&$183.0\pm\phantom01.6$&1.45\\
He I &6678.15&$\phantom{00}38.2\pm\phantom01.2$&$-\phantom089.6\pm\phantom04.9$&$-261.9\pm\phantom03.5$&$194.6\pm\phantom08.0$&1.44\\
\lbrack S II\rbrack&6716.44&$\phantom{000}4.6\pm\phantom00.3$&$-\phantom045.2\pm\phantom08.4$&$-232.2\pm\phantom07.2$&$144.8\pm\phantom07.1$&1.43\\
\lbrack S II\rbrack&6730.82&$\phantom{000}9.9\pm\phantom00.8$&$-\phantom061.5\pm\phantom07.7$&$-239.2\pm19.6$&$214.0\pm29.0$&1.43\\
\lbrack Ar V\rbrack&7005.67&$\phantom{000}1.2\pm\phantom00.3$&$-\phantom089.5\pm22.3$&$-347.6\pm35.3$&$183.7\pm34.3$&1.41\\
\lbrack Ar III\rbrack&7135.79&$\phantom{00}18.9\pm\phantom00.9$&$-\phantom073.8\pm10.4$&$-255.8\pm\phantom05.7$&$187.9\pm10.6$&1.39\\
\lbrack Fe II\rbrack&7155.16&$\phantom{00}43.3\pm\phantom00.9$&$-\phantom085.9\pm\phantom02.9$&$-245.3\pm\phantom03.0$&$170.2\pm\phantom04.2$&1.39\\
\lbrack Fe XI\rbrack&7891.94&$\phantom{000}9.7\pm\phantom00.9$&$-125.9\pm\phantom09.8$&$-336.1\pm27.0$&$218.1\pm25.1$&1.32\\
\lbrack S III\rbrack&9068.6\phantom0&$\phantom{00}15.8\pm\phantom00.9$&$-\phantom053.4\pm\phantom08.3$&$-241.0\pm12.3$&$138.6\pm\phantom06.6$&1.24\\
\lbrack S III\rbrack&9531.10&$\phantom{00}57.4\pm\phantom01.7$&$-\phantom081.6\pm\phantom06.9$&$-267.2\pm\phantom04.5$&$172.3\pm\phantom06.1$&1.22\\
\hline
\label{tab:broadlinesN}
\end{tabular}
\begin{list}{}{}
 \item[$^{\mathrm{a}}$] Fluxes are relative to the flux of H$\beta$ of the unshocked component in the Northern ER: ``100" corresponds
 to $(31.3\pm 0.5) \times 10^{-16}\ergsm$. Fluxes are {\bf not}
 corrected for the extinction. The errors are only statistical, and
  do not include systematic uncertaities (see text).
 \item[$^{\mathrm{b}}$] The velocities are relative to the
  unshocked, narrow component of the emission lines.
 \item[$^{\mathrm{c}}$] The velocity where fluxes are $5 \%$ of the peak flux.
 \item[$^{\mathrm{d}}$] $E(B-V)=0.16{\rm ~mag}$, with $E(B-V)=0.10{\rm ~mag}$ for LMC and $E(B-V)=0.06{\rm ~mag}$ for the Milky Way.
 \end{list}
\end{table*}

\begin{table*}
\centering
\caption{Emission lines from the shocked gas of the southern ER.
}
\begin{tabular}{lclcccc}
\hline
\hline
Emission &Rest wavel.& Relative flux$^{\mathrm{a}}$ &$V_{\rm peak}^{\mathrm{b}}$&$V_{\rm blue}^{\mathrm{c}}$&$V_{\rm red}^{\mathrm{c}}$&Extinction\\
 &$(\ang$)& &($\kms$)&($\kms$)&($\kms$)&correction$^{\mathrm{d}}$\\
\hline
\lbrack Ne V\rbrack&3425.86&$\phantom{00}6.0\pm1.7$&$\phantom09.8\pm28.6$&$-247.6\pm47.2$&$260.7\pm46.1$&2.06\\
\lbrack Ne III\rbrack&3868.75&$\phantom029.5\pm1.5$&$31.7\pm\phantom08.2$&$-173.2\pm\phantom06.7$&$236.4\pm\phantom06.7$&1.94\\
\lbrack S II\rbrack&4068.60&$\phantom041.2\pm2.1$&$31.7\pm\phantom08.2$&$-114.5\pm\phantom06.6$&$192.6\pm\phantom08.1$&1.90\\
\lbrack S II\rbrack&4076.35&$\phantom012.6\pm0.9$&$32.8\pm\phantom09.4$&$-168.2\pm\phantom09.0$&$237.2\pm\phantom09.0$&1.90\\
H$\delta$&4101.73&$\phantom026.5\pm1.9$&$18.8\pm\phantom08.0$&$-118.6\pm\phantom09.1$&$215.2\pm12.1$&1.89\\
H$\gamma$&4340.46&$\phantom057.4\pm2.0$&$28.7\pm\phantom07.2$&$-147.5\pm\phantom05.5$&$210.5\pm\phantom05.9$&1.84\\
\lbrack O III\rbrack&4363.21&$\phantom010.4\pm1.3$&$32.9\pm\phantom08.6$&$-168.9\pm\phantom07.8$&$222.7\pm\phantom07.8$&1.84\\
\lbrack Fe III\rbrack&4658.05&$\phantom{00}3.1\pm0.6$&$12.8\pm18.7$&$-232.3\pm28.6$&$259.6\pm28.1$&1.76\\
He II&4685.7\phantom0&$\phantom010.1\pm1.2$&$39.5\pm15.6$&$-170.9\pm20.0$&$224.2\pm20.0$&1.75\\
H$\beta$&4861.32&$149.8\pm2.9$&$34.1\pm\phantom03.7$&$-175.1\pm\phantom03.5$&$218.5\pm\phantom02.0$&1.71\\
\lbrack O III\rbrack&4958.91&$\phantom018.3\pm0.8$&$14.3\pm\phantom06.2$&$-226.3\pm12.9$&$204.0\pm\phantom06.0$&1.68\\
\lbrack O III\rbrack&5006.84&$\phantom055.3\pm1.5$&$16.4\pm\phantom07.0$&$-214.4\pm\phantom05.8$&$215.6\pm\phantom05.0$&1.67\\
\lbrack O I\rbrack&5577.34&$\phantom{00}3.5\pm0.4$&$38.5\pm\phantom09.3$&$-141.2\pm10.0$&$214.8\pm10.0$&1.57\\
\lbrack N II\rbrack&5754.59&$\phantom063.5\pm1.4$&$40.7\pm\phantom05.9$&$-188.0\pm\phantom03.6$&$207.9\pm\phantom03.0$&1.54\\
He I&5875.63&$\phantom049.7\pm1.2$&$38.2\pm\phantom05.4$&$-215.0\pm\phantom03.0$&$231.6\pm\phantom04.9$&1.53\\
\lbrack O I\rbrack&6300.30&$\phantom069.1\pm1.3$&$32.5\pm\phantom05.1$&$-162.4\pm\phantom01.8$&$193.7\pm\phantom03.8$&1.47\\
\lbrack S III\rbrack&6312.06&$\phantom{00}2.6\pm0.4$&$24.7\pm12.5$&$-158.5\pm16.8$&$213.0\pm16.8$&1.47\\
\lbrack O I\rbrack&6363.78&$\phantom021.7\pm0.8$&$27.1\pm\phantom09.2$&$-143.7\pm\phantom05.4$&$175.1\pm\phantom03.7$&1.47\\
\lbrack N II\rbrack&6548.05&$\phantom024.3\pm0.8$&$31.5\pm\phantom06.0$&$-190.8\pm\phantom04.1$&$196.6\pm\phantom06.6$&1.45\\
H$\alpha$&6562.80&$582.0\pm8.7$&$36.0\pm\phantom02.5$&$-190.9\pm\phantom01.2$&$223.2\pm\phantom00.9$&1.45\\
\lbrack N II\rbrack&6583.45&$\phantom076.1\pm1.4$&$28.8\pm\phantom03.6$&$-184.2\pm\phantom02.9$&$194.2\pm\phantom01.6$&1.45\\
He I&6678.15&$\phantom010.0\pm0.5$&$31.4\pm\phantom08.1$&$-188.4\pm\phantom06.3$&$252.8\pm\phantom06.3$&1.44\\
\lbrack S II\rbrack&6716.44&$\phantom{00}1.3\pm0.2$&$20.1\pm10.0$&$-124.3\pm13.7$&$167.1\pm13.7$&1.43\\
\lbrack S II\rbrack&6730.82&$\phantom{00}3.1\pm0.4$&$27.5\pm10.3$&$-146.0\pm12.4$&$202.4\pm12.4$&1.43\\
\lbrack Ar III\rbrack&7135.79&$\phantom{00}5.9\pm0.6$&$19.6\pm14.5$&$-196.2\pm20.1$&$217.6\pm15.7$&1.39\\
\lbrack Fe II\rbrack&7155.16&$\phantom011.2\pm0.5$&$14.1\pm\phantom06.3
$&$-184.2\pm\phantom09.1$&$199.8\pm\phantom08.4$&1.39\\
\lbrack S III\rbrack&9068.6\phantom0&$\phantom{00}6.8\pm0.8$&$-0.1\pm17.5$&$-272.7\pm24.0$&$240.4\pm42.0$&1.24\\
\lbrack S III\rbrack&9531.10&$\phantom016.5\pm1.4$&$23.4\pm\phantom08.5$&$-275.5\pm25.7$&$216.7\pm12.5$&1.22\\
\hline
\label{tab:broadlinesS}
\end{tabular}
\begin{list}{}{}
 \item[$^{\mathrm{a}}$] Fluxes are relative to the flux of H$\beta$ of the unshocked component in the Northern ER: ``100" corresponds
 to $(31.3\pm 0.5) \times 10^{-16}\ergsm$. Fluxes are {\bf not}
 corrected for the extinction. The errors are only statistical, and
  do not include systematic uncertaities (see text).
 \item[$^{\mathrm{b}}$] The velocities are relative to the
  unshocked, narrow component of the emission lines.
 \item[$^{\mathrm{c}}$] The velocity where fluxes are $5 \%$ of the peak flux.
 \item[$^{\mathrm{d}}$] $E(B-V)=0.16{\rm ~mag}$, with $E(B-V)=0.10{\rm ~mag}$ for LMC and $E(B-V)=0.06{\rm ~mag}$ for the Milky Way.
 \end{list}
\end{table*}


\section{Analysis}

\subsection{The narrow component}
\label{sec:narrowcomp}

\begin{figure*}
\resizebox{\hsize}{!}{\includegraphics{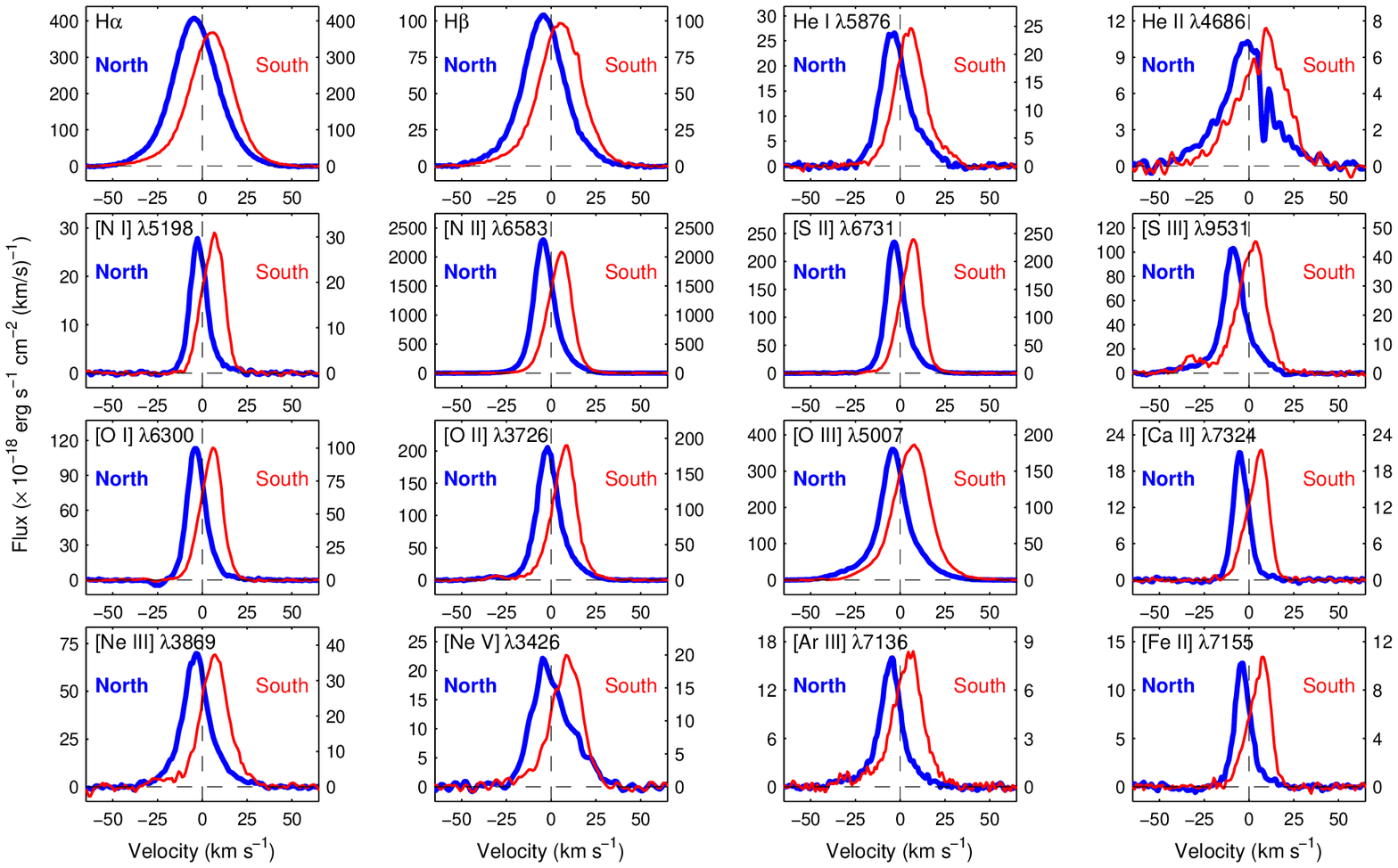}}
\caption{Line profiles of the narrow emission lines from the northern (blue/thick lines)
and southern (red/thin lines) parts of the ER. The zero velocity corresponds to the rest frame of the SN ($286.5\kms$). The left y-axis indicate the flux of the emission lines from the northern part of the ER and the right y-axis indicate the flux from the southern part. The line profiles have been smoothed using a fourth-order Savitzky-Golay filter. Note the thermal broadening of the lighter elements.}
\label{fig:narrowcomps}
\end{figure*}

The dereddened Balmer lines ratios (relative to H$\beta$) are
listed in Table \ref{tab:balmerflux}, together with 
the theoretical values given for case B at $10^4$ K and a gas density
of $10^4{\rm ~cm^{-3}}$
\citep{osterbrock89}. In general, the ratios we find are
close to those predicted by Case B theory, except for our
high $j_{{\rm H}\alpha}/j_{{\rm H}\beta}$ ratios. For both the unshocked and shocked (cf. Sect. \ref{sec:relflux_lineprof}) gas, the H$\beta$ flux appears to be 
stronger relative to the other Balmer lines on the southern side of the ring 
than the northern, maybe by $\sim 10\%$. Given the systematic uncertainty in the relative fluxes, this difference is hardly significant. We note, however, that collisional excitation of H$\alpha$ may be important for the narrow lines \citep[see][]{LF96}.

To estimate the temperature and density of the unshocked ER, we have adopted 
a nebular analysis. For O~II, S~II and S~III we have used a five-level atom, 
and for N~II, O~I and O~III a six-level atom. Atomic data for N~II, O~III, and 
S~II are the same as described in \cite{maran00}, and for O~II we included data
from \citet{osterbrock89} and \citet{MB93}. Data for S~III are from
Hayes (1986), \citet{galavis95} and Tayal (1997).

\begin{figure*}
\resizebox{\hsize}{!}{\includegraphics{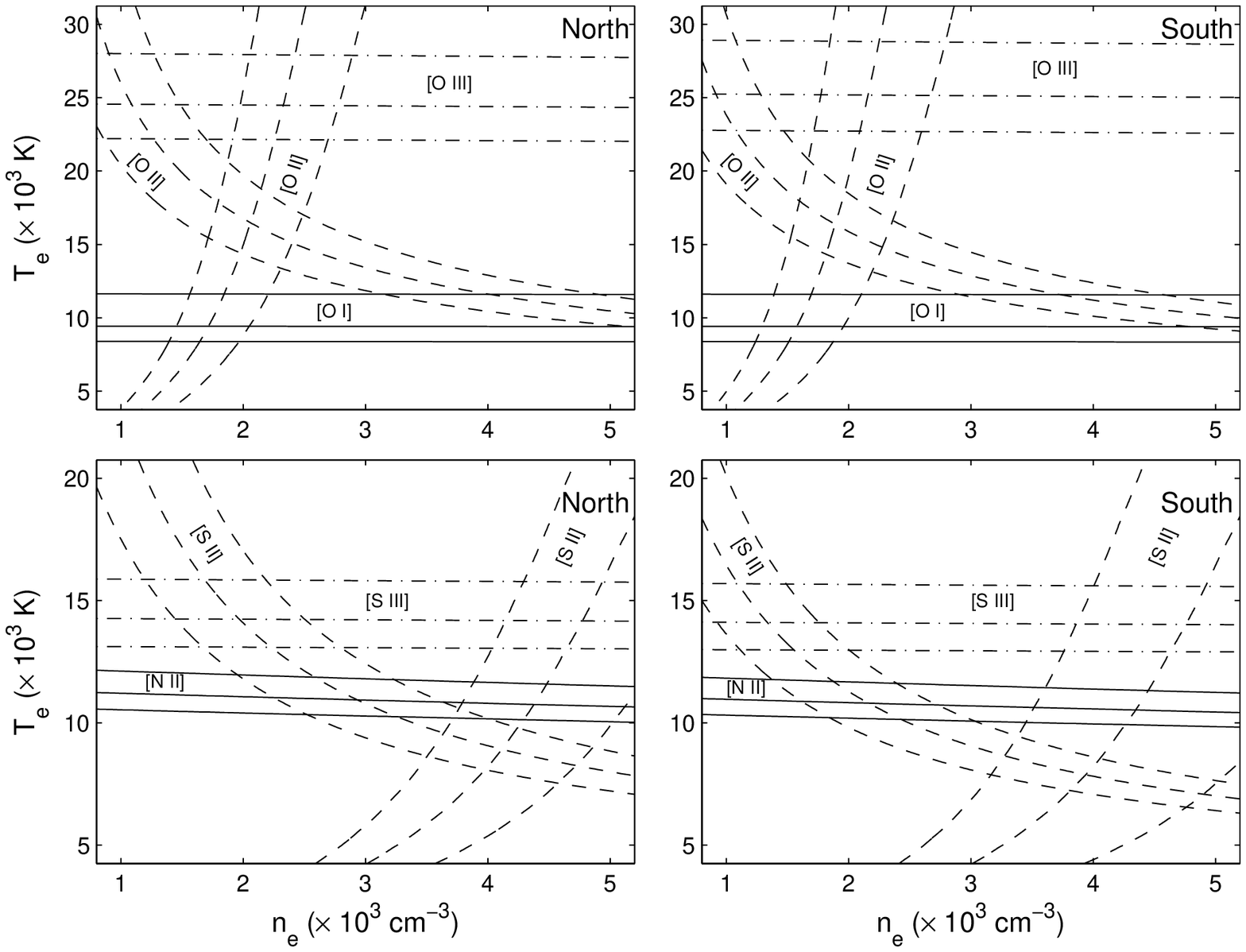}}
\caption{Electron density vs. temperature for the dereddened line ratios
of emission from the unshocked gas. The left panels
show the emission from the northern part of the ring and the right panels
are for the southern part. The line ratios are: [O~I] 
$\lambda\lambda6300,6364/\lambda5577$, 
[O~II] 
($\lambda\lambda7319-7331/\lambda\lambda3726,3729$ and $\lambda3729/\lambda3726$),
[O~III] $\lambda\lambda4959,5007/\lambda4363$, 
[N~II] $\lambda\lambda6548,6583/\lambda5755$, 
[S~II] ($\lambda\lambda4069,4076/\lambda\lambda6716,6731$ and $\lambda6716/\lambda6731$), 
[S~III] ($\lambda\lambda9069,9531/\lambda6312$). 
Each solution curve is embraced by error
curves of at least 15\% to account for the systematic relative flux uncertainties. The exception is for $\lambda3729/\lambda3726$ and $\lambda6716/\lambda6731$ ratios for which the systematic errors should be small. Instead these are embraced by the $1\sigma$ statistical error curves. Note how the rather temperature insensitive ratios $\lambda3729/\lambda3726$ and $\lambda6716/\lambda6731$ together with the temperature sensitive ratios $\lambda\lambda7319-7331/\lambda\lambda3726,3729$ and $\lambda\lambda 4069,4076/\lambda\lambda6716,6731$ can constrain both the density and temperature of [O~II] and [S~II].}
\label{fig:dente_n}
\end{figure*}

In Fig. \ref{fig:dente_n} we show
the allowed temperature-density range for both
parts of the ER, using our six-level atoms for [N~II] and [O~III]. 
As discussed by \citet{maran00}, the narrow emission lines at these
late epochs are expected to mainly come from gas with density significantly
below $10^4 \cc$. Below this density our [O~III] temperatures should be in the 
range $(2.5\pm0.3) \EE4$ K, and the [N~II] temperatures should be confined
to the range $(1.1\pm0.1) \EE4$ K, where the errors stated include 
statistical errors as well as a $\pm10$\% systematic error (see Sect. 2.1).
These results are in general in very good agreement 
with the estimates of \citet{maran00}, who obtained [N~II] temperatures in the
range $9.5\EE3 - 1.3\EE4$ K and [O~III] temperatures in 
excess of $\sim 1.7\EE4$ K, using HST/STIS data from 1998 November 14, with
perhaps a slightly higher [N~II] temperature in the western 
ER ($\mathrm{PA}=283^\circ$) than in the eastern ($\mathrm{PA}=103^\circ$). 
We find no clear evidence for a temperature difference between the
northern and southern parts of the ER.

To get a further handle on the temperature, we can use [O~II]
$\lambda\lambda$3726,~3729 in combination with [O~II]
$\lambda\lambda$7319-7330, as well as [S~II]
$\lambda\lambda$4069,~4076 in combination with [S~II]
$\lambda\lambda$6716,~6731.  The results from our five-level model
atoms are plotted in Fig. \ref{fig:dente_n}.  As the
$j_{\lambda3729}/j_{\lambda3726}$ and
$j_{\lambda6716}/j_{\lambda6731}$ intensity ratios are rather
insensitive to temperature, but the other ratios, i.e.,
$j_{\lambda\lambda7319-7331}/j_{\lambda\lambda3726,3729}$ and
$j_{\lambda\lambda4069,4076}/j_{\lambda\lambda6716,6731}$ are
temperature sensitive, these figures can be used to determine
temperature and density simultaneously for [O~II] and [S~II]. This has
not been done previously for the ER and provides a more accurate determination as input for models such as those of \citet{LF96}. For
[O~II] there is an overlap in temperature ($(1.7\pm0.5)\EE4$~K) as
well as in electron density, $(2.0\pm0.5)\EE3 \cc$, of the two parts of
the ER. For [S~II], the overall preferred range of electron density
vs. temperature is in this case $(4\pm1)\EE3 \cc$ and
$(6.5-11)\EE3$ K. The [S~II] temperature could be slightly lower on the
southern side than in the northern.

Again, we may compare our 
results against those of \citet{maran00} who estimated the electron density 
from [S~II]. They obtained a value of $\sim 4\EE3 \cc$ for both the western
and eastern parts of the ER, i.e., very similar to ours, albeit for
other parts of the ring. They did not estimate any value for the
[O~II] density. Our nebular analysis is completed by temperature
estimates using [O~I] and [S~III] for which we find $(10\pm2)\EE3$
K and $(1.3-1.6)\EE4$ K, respectively.

In summary, we estimate temperatures in the range between $\sim
6.5\EE3$ and up to at least $2.2\EE4$ K. Listed in order of increasing
temperature for the ions we have analyzed, we find S~II, O~I, N~II,
S~III, O~II and O~III. There is a hint of a somewhat lower temperature
on the southern side than the northern, especially for [S~II] and
maybe also for the Balmer lines. In a similar way we estimate
densities in the range between $\sim 1.5\EE3 \cc$ and $\sim 5.0\EE3
\cc$ based on [O~II] and [S~II], respectively. This probably also
brackets the electron densities for the other ions
(Fig. \ref{fig:dente_n}). The exception is the [O~III] emission, which
models show mainly to come from gas with an electron density which is
lower than $\sim 1.5\EE3 \cc$ \citep[e.g.][]{LF96}. These models show
that the gas with low density remain in
higher ionization stages longer.  There is no obvious difference in
density between the two sides of the ring.

\begin{table*}
\centering
\caption{Derived temperatures from the widths of the narrow lines using 
[Ar~III] $\lambda$7136 and [Fe~II] $\lambda$7155 as templates for the 
instrumental broadening and the expansion of the ring.}

\begin{tabular}{l c c c c}
\hline
\hline
Template: &[Ar~III] $\lambda$7136& [Ar~III] $\lambda$7136 & [Fe~II] $\lambda$7155&  [Fe~II] $\lambda$7155\\
Position: & north & south & north & south \\
Emission line & $T$ ($10^3$ K) & $T$ ($10^3$ K) & $T$ ($10^3$ K) & $T$ ($10^3$ K) \\
\hline
H$\alpha$&$10.5\pm\phantom00.5$&$\phantom08.0\pm\phantom00.5$&$14.6\pm0.6$&$\phantom012.4\pm\phantom00.6$\\
H$\beta$&$\phantom08.0\pm\phantom00.5$&$\phantom06.9\pm\phantom00.6$&$11.6\pm0.6$&$\phantom011.2\pm\phantom00.7$\\
H$\gamma$&$\phantom08.0\pm\phantom01.0$&$\phantom06.5\pm\phantom00.5$&$12.0\pm1.0$&$\phantom010.7\pm\phantom00.9$\\
H$\delta$&$\phantom09.0\pm\phantom01.0$&$\phantom04.3\pm\phantom01.3$&$13.0\pm1.0$&$\phantom{00}8.0\pm\phantom01.0$\\

He I $\lambda$5875.7&$\phantom07.0\pm\phantom01.0$&-&$13.3\pm1.8$&$\phantom011.5\pm\phantom01.5$\\

He I $\lambda$6678.2&$\phantom05.5\pm\phantom01.5$&-&$13.8\pm1.8$&$\phantom010.8\pm\phantom01.8$\\

He II $\lambda$4685.7&$28.0\pm\phantom08.0$&$25.0\pm10.0$&$40.0\pm5.0$&$\phantom040.0\pm10.0$\\

[N II]~$\lambda$5754.5&-&-&$22.0\pm2.0$&$\phantom020.0\pm\phantom02.0$\\

[N II]~$\lambda$6547.9&-&-&$16.3\pm1.8$&$\phantom{00}8.8\pm\phantom01.8$\\

[N II]~$\lambda$6583.3&-&-&$14.5\pm1.5$&$\phantom{00}9.5\pm\phantom01.5$\\

[O I]~$\lambda$6300.3&-&-&$10.0\pm1.5$&$\phantom{00}3.5\pm\phantom01.5$\\

[O I]~$\lambda$6363.8&-&-&$\phantom09.0\pm1.5$&$\phantom{00}6.0\pm\phantom03.0$\\
 
[O II]~$\lambda$3726.0&-&-&$25.0\pm3.0$&$\phantom{00}9.0\pm\phantom02.0$\\

[O II]~$\lambda$3728.8&-&-&$29.0\pm3.0$&$\phantom011.0\pm\phantom02.0$\\
 
[O III]~$\lambda$4363.2&$27.0\pm\phantom07.0$&$55.0\pm15.0$&$51.3\pm6.3$&$110.0\pm20.0$\\

[O III]~$\lambda$4958.9&$17.0\pm\phantom03.0$&$49.0\pm\phantom06.0$&$35.0\pm3.0$&$105.0\pm10.0$\\

[O III]~$\lambda$5006.8&$17.0\pm\phantom03.0$&$43.0\pm\phantom08.0$&$37.0\pm3.0$&$110.0\pm10.0$\\
 
[Ne III]~$\lambda$3868.8&$23.0\pm\phantom03.0$&-&-&-\\
 
[Ne V]~$\lambda$3425.9&$40.0\pm10.0$&$15.0\pm10.0$&-&-\\
 
\hline
\label{tab:ringtemp}
\end{tabular}
\end{table*}

Another way of deriving the temperature of the emitting gas is from the 
width of the line profiles, where the FWHM line width due to thermal 
broadening relates to the temperature as

\begin{equation}
V_{\rm FWHM} = 21.40\left({{T\over 10^4A}}\right)^{1/2} \kms
\label{eq:therm}
\end{equation}

\noindent Here $A$ is the atomic weight of the emitting ion. Because the ER 
is an extended source, the UVES slit samples different parts of it, 
each of which having its own macroscopic and thermal broadening. 
Equation (\ref{eq:therm})
can therefore only provide an upper limit to the actual temperature.
Because 
the thermal contribution
decreases with increasing ion mass,
the line profiles of the heaviest ions should have minimum thermal
contribution. We have chosen two such lines, but with different level
of ionization to test the importance of spatial structure (cf. above), 
namely [Ar~III] $\lambda$7136 and [Fe~II] $\lambda$7155.
These lines also have the advantage of both being in the red part of the spectrum where the velocity
resolution of the profiles are better than in the blue.
To estimate the thermal contribution of the other lines
we assumed that their macroscopic distributions over the ring are
similar to that of the template line. The thermal component is then the
deconvolution of the total line profile and the template.
Table \ref{tab:ringtemp} shows the estimated
temperatures for some lines.

\begin{figure}
\resizebox{\hsize}{!}{\includegraphics{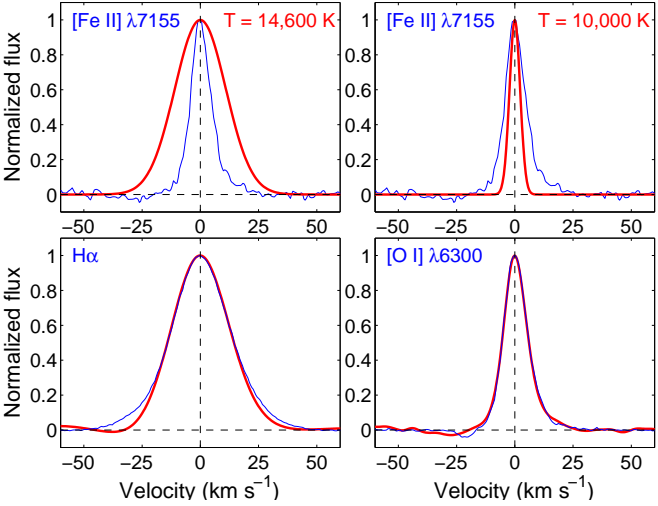}}
\caption{The upper left panel shows [Fe II] $\lambda$7155 north
(blue/thin line) and a thermal (gaussian) line profile for hydrogen with a
temperature of $14.6\times 10^3$ K (red/thick line). The lower left panel
shows the result after convolving [Fe II] with the thermal component
(red/thick line) vs. the observed H$\alpha$ line profile at the
corresponding position (blue/thin line). The right panels show
the same for [O I] $\lambda$6300, and a thermal line profile
with a temperature of $10\times 10^3$ K for oxygen. The zero velocity corresponds to the peak flux of the lines.}
\label{fig:FeII_conv}
\end{figure}

Due to their low atomic mass, lines of H and He are dominated by
thermal broadening and the spatial structure is less important. The
opposite is true for lines emitted by heavy ions such as sulphur. This
is exemplified in Fig. \ref{fig:FeII_conv} for H$\alpha$ and [O~I]
$\lambda$6300, where we in both cases have used the [Fe~II] line as
template. The observed profiles and fits to these are shown in the
lower panels. In the upper panels we show the template line together with
the thermal component of H$\alpha$ (top) and [O~I] $\lambda$6300
(bottom) needed to get a good fit. It is obvious that the error
in estimated temperature becomes significant when the thermal width is
smaller than the width of the template. As a matter of fact, the
analysis becomes pointless for ions heavier than neon, and we have
omitted such ions from Table \ref{tab:ringtemp}.

Table \ref{tab:ringtemp} shows that the line profiles indicate
hydrogen temperatures of $\sim 10^4$ K, with H$\alpha$ giving the
highest temperature, regardless of template ion used and position of
the ring. This is interesting since a temperature slightly in
excess of $10^4$ K is required for collisional excitation to come into
play, as hinted by the Balmer line ratios. The fact that the unshocked
ring is supposed to have a high abundance of only partially ionized
hydrogen at these late stages \citep[cf.][]{LF96}, of course also boosts
collisional excitation relative to recombination.

For He I we obtain
similar temperatures as for the Balmer lines, whereas He II
temperatures are in the range $(2.5\pm1.0)\times10^4$ K with
[Ar~III] as template and about $(4.0\pm1.0)\times10^4$ K with [Fe~II] 
as template. It appears likely that [Fe~II] is a better template
for H I and He I, whereas [Ar~III], with its higher ionization potential,
should be better for He II. There is an indication that the temperature is slightly higher
on the northern side for the gas emitting H and He lines, which is 
consistent with the Balmer line ratios (see Table \ref{tab:balmerflux}). 
It is tempting to suggest
that this higher temperature is a result of higher shock
activity on that side of the ring, which could photoelectrically excite
the unshocked gas.

For the metal lines the thermal broadening is much smaller than for H and He
and the method becomes less exact. Nevertheless, it can pick up trends
in the estimated temperature, and it offers an independent test of line ratio
temperatures. Such a trend is seen for the lightest metal ion N~II for 
which the northern side gives a higher temperature than the southern, i.e.,
in agreement with line ratio temperatures.  It is also interesting that 
[N~II]~$\lambda5755$ gives a higher temperature than 
the $\lambda\lambda6548,6583$ doublet. The latter comes from a lower 
excitation level and is biased to come from regions of lower temperature and 
density. The same is seen for the [O~III] counterpart [O~III]~$\lambda4363$
compared to [O~III]~$\lambda\lambda4959,5007$. Whereas the [N~II] temperatures
are in fair agreement with those from line ratios, the [O~III] temperatures
are inconsistent with the line ratio temperature for the southern side.
This is most likely due to the fact that the [O~III] lines are much stronger
on the northern side and that there could be ``leakage" of emission from
the northern side to the blue wing of the emission from the southern side.
The southern side line profile thus becomes too broad and the estimated temperature
too high. The same is true also for the [Ne~III] lines.
Table \ref{tab:ringtemp} clearly shows the importance of using the correct
template line for [O~III], and we have retained the figures in the
last two columns only to highlight these effects. The only [O~III] line
profile temperatures to be trusted are for [Ar~III] as template and
for the northern side. Those temperatures are also in general agreement
with line ratio temperatures.

For the other metal lines we note that
the [Ne~III]~$\lambda3869$ agrees with the [O~III] temperatures, as expected,
and that [Ne~V] indicates a higher temperature. The uncertainty of the
latter is, however, very large due the noisiness of that line.
For [O~I] and [O~II] we see the same effect as for [N~II], i.e., the
northern side appears hotter. The suggested temperature difference between the two sides
of the ring is, however, too pronounced compared to line ratio temperatures 
for [O~II], even if one would consider a difference in density of the 
[O~II]-emitting gas on the two sides of the ring. It should be kept in mind
that [O~II]~$\lambda\lambda3726,3729$ lie far out in the blue where the 
spectral resolution is worse and the line profile temperatures less accurate.
In addition, the line ratio temperature included the [O~II] lines between 
7319-7321 \AA\ which also affects a direct comparison between line ratio
and line profile temperatures for [O~II].
 
\subsection{The intermediate component}
In Tables \ref{tab:broadlinesN} and \ref{tab:broadlinesS} we give
fluxes of the intermediate line components. The table also provides
velocities of the peak of the line profiles, as well as maximum red
and blue velocities of the lines. All velocities are with respect to
the corresponding narrow line component.

\begin{figure*}
\resizebox{\hsize}{!}{\includegraphics{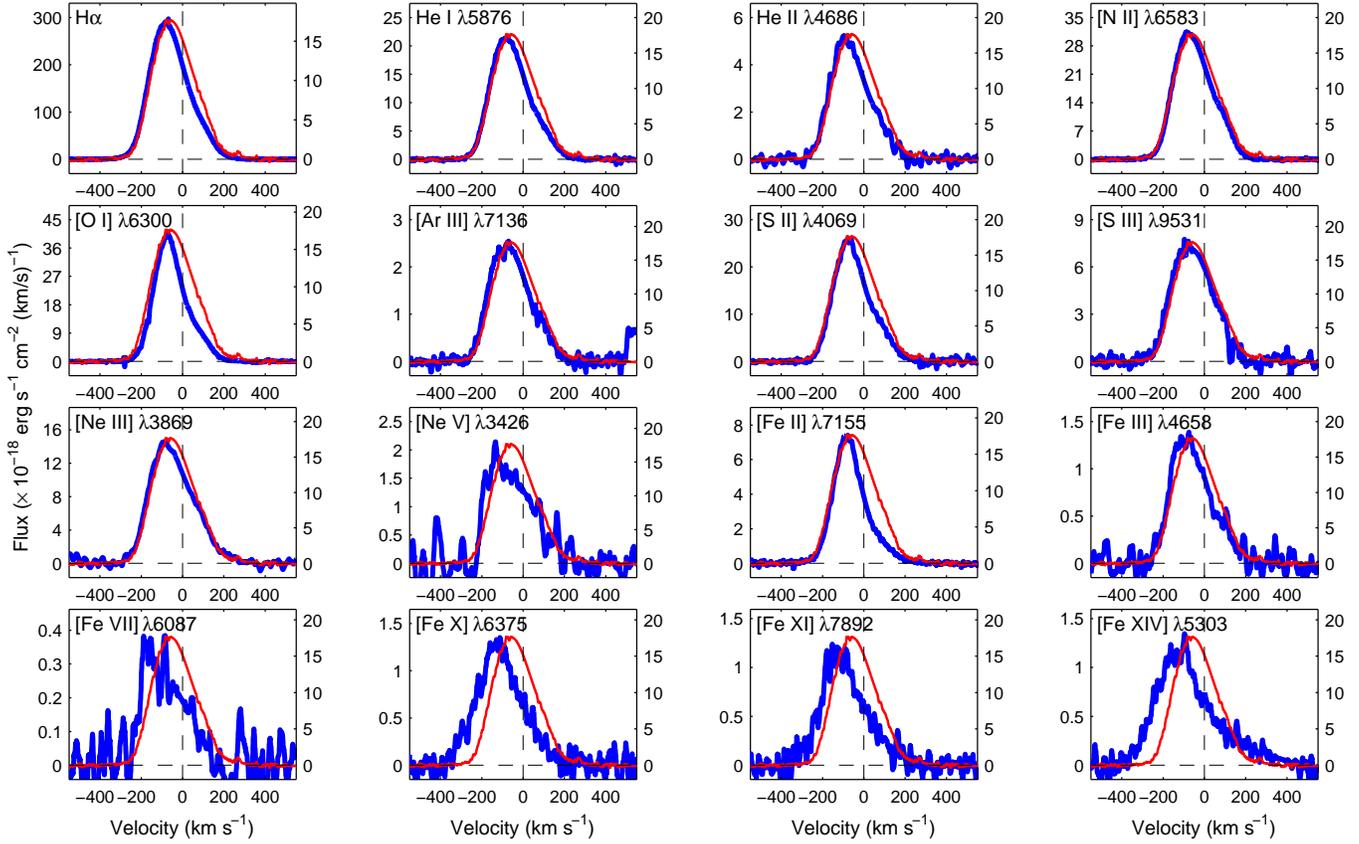}} 
\caption{Line profiles of the broad emission lines (blue/thick lines) vs. [O~III] $\lambda5007$ (red/thin lines) from the northern part of the ER. The zero velocities are relative to the peak velocities of the unshocked, narrow component of the emission lines. The flux of [O III] is indicated by the right y-axis. The line profiles have been smoothed using a fourth-order Savitzky-Golay filter. Note the difference in profiles between the lower ionization lines and the coronal lines.}
\label{fig:lineprofiles_0210n}
\end{figure*}

\begin{figure*}
\resizebox{\hsize}{!}{\includegraphics{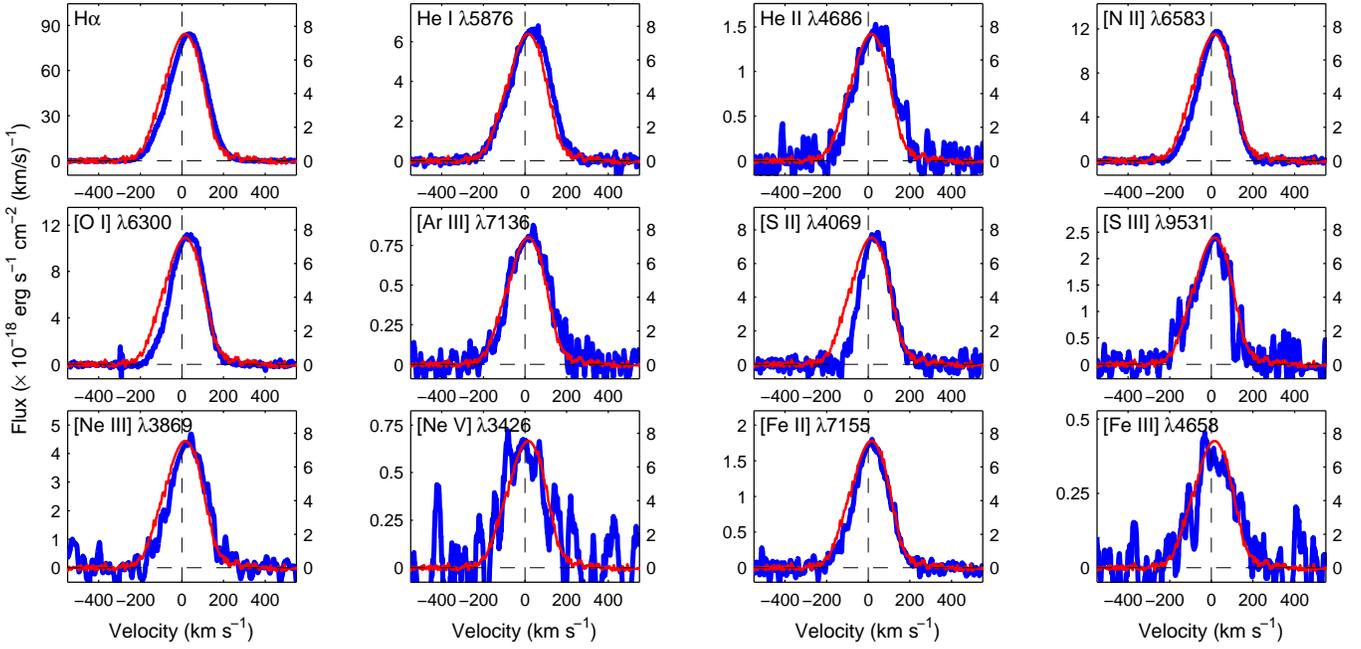}} 
\caption{Line profiles of the broad emission lines (blue/thick lines) vs. [O~III] $\lambda5007$ (red/thin lines) from the southern part of the ER. The zero velocities are relative to the peak velocities of the unshocked, narrow component of the emission lines. The flux of [O III] is indicated by the right y-axis. The line profiles have been smoothed using a fourth-order Savitzky-Golay filter. Because of the low flux we do not show the coronal lines for this region.}
\label{fig:lineprofiles_0210s}
\end{figure*}

In Figs. \ref{fig:lineprofiles_0210n} and \ref{fig:lineprofiles_0210s}
we show a selection of line profiles of different ionization stages
from the intermediate component. To facilitate a comparison with other
lines we have in each figure also included the line profile of the
high S/N line of [O III]~$\lambda$5006.8. From the figures it is
apparent that the S/N varies by a large factor from the strongest
lines to the weaker, which have fluxes less than a percent of the
strongest. We also see that both the profile and the peak velocities
differ strongly between the northern and southern components (see Sect. \ref{sec:lprof}). In
general, the northern component has the highest flux, and we will
therefore in the following concentrate our discussion on this part.

\begin{table}
\centering
\caption{The dereddened Balmer line fluxes relative to H$\beta$ for the shocked gas. Only the $1\sigma$ statistical errors are given. Systematic errors are discussed in the text.}
\begin{tabular}{c c c c c}
\hline
\hline
Position & H$\alpha$ & H$\beta$  & H$\gamma$ & H$\delta$ \\
\hline
north & $342.3\pm0.4$ & $100$ & $45.13\pm0.09$ & $22.71\pm0.07$\\
south & $328.6\pm0.5$ & $100$ & $41.4\pm0.2$ & $19.0\pm0.2$\\
Case B& $287$ & $100$ & $46.6$ & $25.6$\\
\hline
\label{tab:balmerflux_sh}
\end{tabular}
\end{table}

\subsubsection{Nebular diagnostics}
\label{sec:relflux_lineprof}

As for the narrow lines in Sect. \ref{sec:narrowcomp}, we have used the fluxes in
Tables \ref{tab:broadlinesN} and \ref{tab:broadlinesS} to probe the
temperatures and the electron densities of the shocked gas. The
results from our multilevel model atoms are plotted in
Fig. \ref{fig:dente_b}. The figure panels also include shaded areas
which encapsulate likely isobars (expressed in pressure units of
$10^{10}$~K~cm$^{-3}$). It is, as we will discuss elsewhere, a good
assumption to assume constant pressure for the radiative shocks
propagating into the ER. The reason we find these isobars more likely
than others is that an [O~III] temperature below the $\sim 10^4$~K
typical of photoionized gas seems unlikely (marking the right boundary
of shaded are), and that the [O~III] temperature should be higher than
the temperatures of [N~II] and [S~II], which roughly coincides with an
[O~III] temperature of $\sim 4\EE4$~K (giving the left boundary). The
latter temperature is more typical of shock excitation (see next section). From Fig. \ref{fig:dente_b} this temperature interval
for [O~III] constrains the electron density in the [O~III] emitting
region to be between $10^6 - 10^7 \cc$ for both the northern
and southern parts of the ER.

Furthermore, Fig. \ref{fig:dente_b}
shows a very reasonable temperature structure of the emitting gas in
the shaded area, with the temperature decreasing, in order, through
the [O~III], [N~II]/[S~III], [O~I], [S~II] and [N~I] regions. In the
[N~I] region it could be as low as $(4-5)\EE3$~K, although the exact
temperature depends on the degree of ionization there.  For the
unshocked ER we found the sequence [O~III], [S~III], [N~II], [S~II]
and [O~I], which would be the same also for the shocked gas if we
limit ourselves to the upper densities of the shaded area in
Fig. \ref{fig:dente_b} where the [S~III] temperature is higher than
that of [N~II]. However, this can only be regarded as a hint since
there is no clear reason why the temperature stratification should be
exactly the same in the unshocked and shocked gas.

In general our results agree with the nebular analysis performed by
\citet{pun02}, as they found roughly the same (within errors) elemental
sequence of decreasing temperature. However, their data suffer from
comparatively large uncertainties for some of the line ratios, making a
direct comparison rather difficult. Nevertheless, we find good
agreements for the [N II] and [O III] densities and temperatures. From
our nebular analysis we find, however, consistently higher
temperatures for [O~I] than those estimated by \citet{pun02}, although
the flux ratios are similar. The origin of this is not clear.

\subsubsection{Origin of the optical lines}

A detailed discussion of the characteristics of the emission from the
shocked gas can be found in \citet{pun02}. Here we only discuss the
points relevant to our observations.  

It is easy to exclude thermal
broadening as the cause of the intermediate components. A $V_{\rm
FWHM} \sim 200 \kms$ would indicate a temperature of $T \sim 10^6$ K,
and at this temperature the observed ions would be collisionally
ionized. Instead, these optical emission lines of the low ionization
species originate from the photoionization zone behind the radiative
shock, with a temperature of $\sim 10^4$ K \citep{pun02}. It is
therefore obvious that the shape of the line profiles must be
dominated by the shock dynamics, rather than thermal broadening.

We have calculated models of radiative shocks for several shock
velocities between 100 -- 500 $\kms$, using a shock code based on the
most recent atomic data.  The details of this, as well as models for
the shock emission in SN 1987A, will be discussed in
\citet{fra07}. Here we only make some general remarks about the origin
of the different lines. As an example, we take the $V_s = 300 \kms$
shock model with a pre-shock density of $10^4 \cc$, similar to that of
the ER.

Immediately behind the shock the temperature for this shock
velocity is $\sim 1.5 \times 10^ 6$ K. This is where part of the soft
X-ray lines \citep{zhe05,zhe06} and the coronal lines discussed in
\citet{gro06} arise. For shock velocities in the range $V_s = 250 -
400 \kms$ the thickness of this collisionally ionized, hot region is
\begin{equation}
D_{cool} \approx 8.5 \times 10^{14} \left({n_{0} \over 10^4 \cc}\right)^{-1} \left({V_s \over 300
  \kms}\right)^{4.4}  \rm \ cm,
\label{eq:cooldist}
\end{equation}
where $n_{0}$ is the electron density in the pre-shock gas, i.e., in
the ER. At this point the temperature is $\sim 10^ 5$ K. The gas then
undergoes a thermal instability and cools to $\sim 10^ 4$ K in a very
thin region. After this the gas temperature is stabilized by
photoionization heating, and a partially ionized zone takes over. In
this zone the degree of ionization, $x_e$, and temperature decrease
from $x_e \sim 1.0$ and $T_e \sim 10^4$ K, respectively, to very low
values.

The models also show that the time for a shock to become
radiative, $t_{\rm cool}$, is
\begin{equation}
t_{\rm cool} \approx 6.7 \left({n_0 \over 10^4 \cc}\right)^{-1} \left({V_s\over 300\kms}\right)^{3.4} \rm years.
\label{eq:cooltime}
\end{equation}
\noindent

In the absence of a magnetic field the density in the photoionized
region is
\begin{equation}
n_{\rm ph} \approx 10^2 \left({V_s \over 100 \kms}\right)^2
\left({T_{\rm ph} \over 10^4 ~\rm K}\right)^{-1} n_0 \ ,
\label{eq:preshdens}
\end{equation}
where $T_{\rm ph}$ is the temperature in the photoionized
zone. If there is a sufficiently strong magnetic field this may limit
the compression behind the shock to $\sim (8 \pi \rho_0)^{1/2}
V_s/B_0$, where $\rho_0$ and $B_0$ are the density and magnetic field
in the pre-shock gas, respectively \citep{cox72}. For $V_s \sim 300
\kms$ and $n_{0} \sim 10^4 \cc$ we therefore expect a density of
$\sim 10^7 \cc$, which is in good agreement with the density we infer
in this region, $4 \EE6 - 2 \EE7 \cc$
(Sect. \ref{sec:relflux_lineprof}). There is therefore not from this
observation any need for a dominant magnetic field.

Besides the coronal lines, also the [Ne V] lines come from the
cooling, collisionally ionized region with $(1-3) \times 10^5$ K. In
terms of emissivity, [Ne V] takes over roughly where [Fe X] drops. The
[O III] and [Ne III] lines arise from a wide range of temperatures,
$(2-10) \times 10^4$ K, in the collisionally ionized zone, but also
have a contribution from the photoionized zone at $\sim 10^ 4$ K.

Both the H I and He I lines are dominated by the photoionized region
just behind the cooling region at $\sim 10^ 4$ K. Most of the emission
comes from the region where $x_e \gsim 0.1$, with a thickness of only
$\sim 10^{12}$ cm.  The low ionization lines from [O I], [N II], [S
II], and [Fe II] all come from the photoionized zone, but the emission
extends to a lower temperature and ionization compared to the H I and
He I lines. With the exception of the coronal lines, the optical line
emission therefore require radiative shocks.

\subsubsection{Line profiles}
\label{sec:lprof}

The interpretation of the shape of the line profile is not
straightforward, although it is clear that it is dominated by the
fluid motion behind the shocks, while thermal broadening should only
have a minor influence. As discussed in Sect.~\ref{sec:introd}, it is,
however, likely that there is a range of shock velocities, determined
by the blast-wave velocity, density and the geometry of the blobs. 

Furthermore, the line profiles only reflect the projected shock velocities
along the line of sight, and consequently, the FWZI velocity is only a
lower limit to the maximum velocities of the radiative shocks driven
into the protrusions. At this epoch (2002 October) the shocks have had about $\sim5$ years to cool since first impact, and from the FWZI of the line profile
of H${\alpha}$ (see Tables \ref{tab:broadlinesN} and
\ref{tab:broadlinesS}), we estimate that shocks with velocities $\lsim
260 \kms$ have had enough time to cool. This velocity
depends on the density, $V_{\rm cool} \propto n_0^{-0.29}$. As
faster shocks become radiative the maximum shock velocity resulting in
radiative shocks is expected to increase with time. Therefore, the
widths of the emission lines are also expected to increase. 

If we compare the line profiles of the different lines we find some
very interesting differences. The low ionization ions such as H$\alpha$, He
I, [O III] and up to [Ne V] and [Fe VII], all have very similar
profiles, with peak velocities $-60 {\rm ~to~} -80 \kms$, with $V_{\rm
blue}\sim -260 \kms$ for the northern part of the ER
(Fig. \ref{fig:lineprofiles_0210n}). The coronal lines have, however,
a considerably higher peak velocity of $\sim -130 \kms$, and extend
for [Fe XIV] to $V_{\rm blue} \sim -387 \kms$. This indicates that
these two groups of lines do not arise from exactly the same regions.

An advantage with this comparatively early epoch is that the emission
from the northern part of the ER is still dominated by Spot 1. At
more recent epochs, a large number of blobs along the ER, all
with different line of sight projections, contribute to the
emission. There should therefore at the epoch discussed in this paper
not be any serious contamination from other spots with different line-of-sight angles and shock velocities. The line profile is therefore
dominated by the geometry and density distribution around the
blob. Further support for this comes from more recent observations in
2005 with the SINFONI instrument at VLT \citep{kjaer07}. These
adaptive optics observations have lower spectral resolution, but
higher spatial resolution than UVES, $\sim 0\farcs2$. These
observations show that even in 2005, when many blobs contribute
to the emission in the ER, there is an agreement between the peak
velocity of the lines in the direction of the UVES slit, as measured
by SINFONI, and that obtained from the UVES observations. In contrast to our UVES observations, the SINFONI observations, do not show the full distribution of velocities, only the average at each point.

In this context we note the difference between the northern and
southern line profiles shown in Figs. 9 and 10 (see also table 4 and
5). While the peak velocities, $V_{peak}$, are $-50-(-90)\kms$ for the
low and intermediate ionization lines from the northern part of the
ER, they are in the range $10-40\kms$ for the southern part. As
discussed in Kjaer et al., these differences are consistent with the
general geometry of the ER. Because of the low fluxes we can, however, not say much about the profiles of the coronal lines from the southern part.

When we compare the line profiles of the low ionization lines (e.g. [O~I], [S~II] and [Fe~II]) with the intermediate ionization lines (e.g. [O~III] and [S~III]) we find that for the northern part the blue wings are very similiar. However, the red wings of the low ionization lines are significantly weaker compared to the intermediate ionization lines. The opposite is true for the southern part. This indicates that the excitation conditions are different in the two regions.

\citet{pun02} find that a simple spherical geometry well explains
the basic features of the line profile from Spot 1. They, however,
find it difficult to fit the wing with velocities $\la -150\kms$ with
this model, and therefore invoke a 'turbulent' smoothing, possibly
originating from instabilities in the shock structure. While this may
be reasonable, it is difficult to understand how this can produce
velocities higher than the shock velocity. Instead, we believe that it
is more reasonable that the wings come from the highest velocity
shocks, where only a small fraction of the gas, presumably that with
the highest densities, has had time to cool. There should then also be
a contribution from adiabatic shocks, with cooling time longer than
that indicated from Eq. (\ref{eq:cooltime}). \citet{pun02} suggest that
these could be responsible for some of the X-ray emission observed.

This picture is supported by our observations of the larger extent of
the coronal lines, up to $\sim 400 \kms$, compared to the
lower ionization lines. Some of the coronal emission may therefore
originate from adiabatic shocks, which are also likely to give rise to
some of the soft X-ray emission seen by Chandra
\citep{zhe05,zhe06}. If this is correct one expects the maximum
velocity of both the coronal lines, and especially the low ionization
lines, to increase with time. 

We note that the shock velocity should be $\sim 4/3 \times$ the gas
velocity immediately behind the shock, as reflected in the
line widths of the coronal lines. The velocity of the cool gas seen in
the low ionization lines should, on the other hand, be close to that
of the shock velocity. Therefore an extent of $\gsim 400 \kms$ of the
coronal lines indicates a shock velocity of at least $\gsim 510 \kms$. 

\begin{figure}
\resizebox{\hsize}{!}{\includegraphics{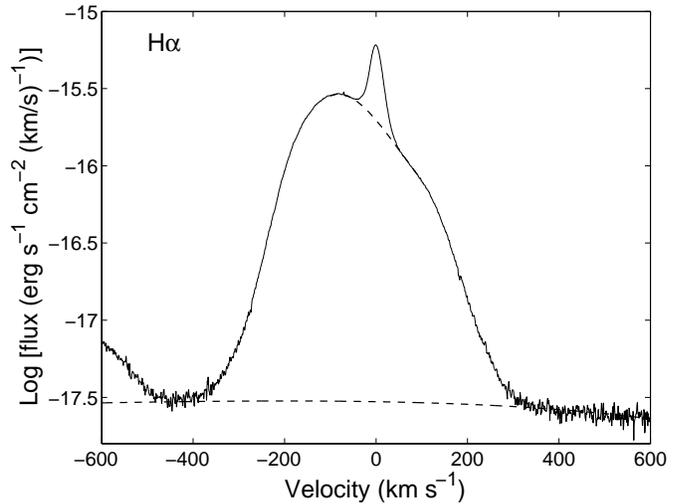}} 
\caption{The narrow and intermediate components of H$\alpha$ for the northern part
of the ER.  The dashed lines show the estimated zero intensity levels
for the narrow and intermediate components respectively. The blue wing
of the intermediate component is only barely separated from the
emission of [N II] $\lambda6548$. The zero velocity corresponds to the peak flux of the narrow component.}
\label{fig:Ha_profile}
\end{figure}

\begin{figure*}
\resizebox{\hsize}{!}{\includegraphics{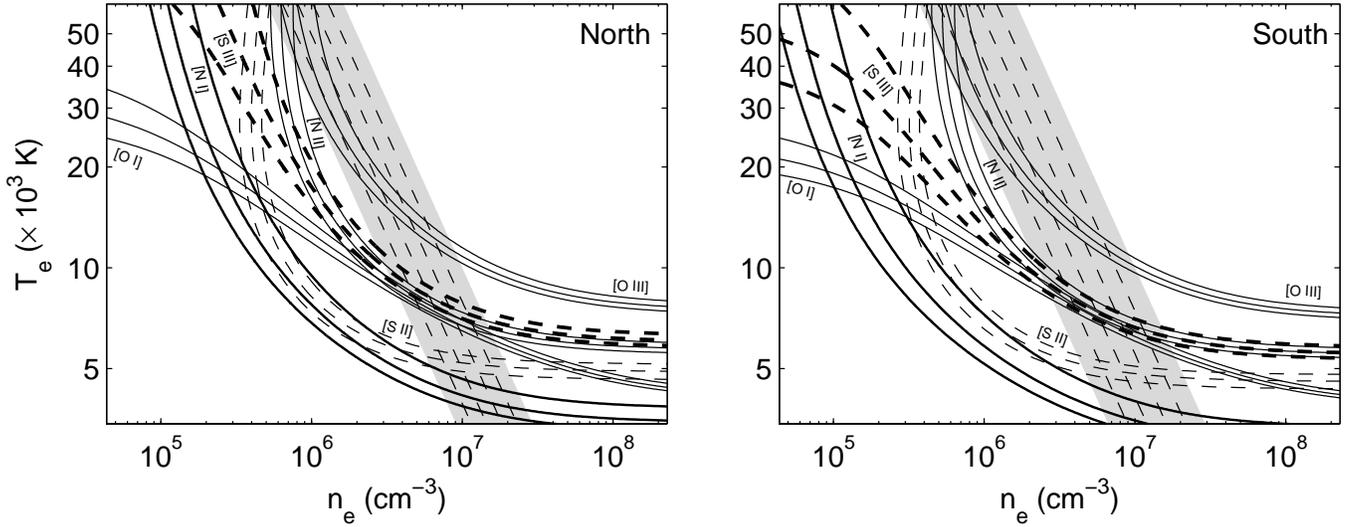}}
\caption{Electron density vs. temperature for the dereddened line ratios
of emission from the shocked gas. The left panel
shows the emission from the northern part of the ER and the right panel
is for the southern part. The line ratios are: [O~I] $\lambda\lambda6300,6364/\lambda5577$, 
[O~III] $\lambda\lambda4959,5007/\lambda4363$, 
[N~I] $\lambda\lambda5198,5200/\lambda3467$, 
[N~II] $\lambda\lambda6548,6583/\lambda5755$, 
[S~II] $\lambda\lambda4069,4076/\lambda\lambda6716,6731$, 
[S~III] $\lambda\lambda9069,9531/\lambda6312$.
Each solution curve is embraced by error
curves of at least 15\% to account for the systematic relative flux uncertainties. The shaded region marks the range for 
possible isobaric solutions. Four isobars are plotted (dashed lines) 
and the pressures are (in units of $10^{10}\mbox{ K}\cc$): $3.9(3.1)$,
$5.0(4.1)$, $6.3(5.5)$ and $7.9(7.4)$. 
The values in brackets are for the southern part.}
\label{fig:dente_b}
\end{figure*}

Although the emission from the coronal lines clearly comes from higher
velocities than the lower ionization stages, it is important to
realize that there are also in these evidence for gas at
velocities considerably higher than $V_{\rm blue}$ or $V_{\rm red}$,
which only gave the extension to 5\% of the peak flux. In particular,
the H$\alpha$ line, with the best S/N of all lines, shows clear
evidence of an extension to considerably higher velocity than $V_{\rm
blue} \approx -260 \kms$. To illustrate this we show in
Fig. \ref{fig:Ha_profile} H$\alpha$ on a logarithmic flux scale,
From this we see that this line has a weak blue extension up to $\sim
-450 \kms$, while the red extension is up to $\sim 350 \kms$.
Therefore, at least a small fraction of these high velocity shocks
have time to cool. Most likely this comes from shocked gas with
densities considerably higher than that typical in the ER, $\sim 10^4
\cc$. It is in this connection interesting that \citet{LF96} find
evidence for densities $\gsim 3\EE4 \cc$ in the ER. From
Eq. (\ref{eq:cooltime}) we find that for a shock velocity of $450
\kms$ gas with a density of $\gsim 4\EE4 \cc$ should have time to
cool.

We finally remark that it is well known that radiative shocks in
this velocity range are unstable to instabilities with wavelength of the order
of the cooling length \citep{racim82,stri95,suth03}. These may add
further to the already complex line profiles.

\section{Summary}

Our UVES observations show the dynamics of the shock interaction in SN
1987A with unprecedented spectral resolution. We find evidence for
shock velocities up to $\sim 500 \kms$. From the larger extent
of the line profiles of the coronal [Fe X-XIV] lines compared to the
low ionization lines, we argue that the highest velocities come from
adiabatic shocks, which have not had time yet to cool. Shocks with
velocity less than $\sim 250 \kms$ are radiative and give rise to the
rest of the optical lines. We do, however, also find evidence for a
minor component of radiative shocks with higher velocity. As more and
more gas cools we expect the width of especially the low ionization
lines to increase.

While the coronal lines and high ionization lines like [O III] and
[Ne III-V] arise in the collisionally ionized hot gas behind the
shocks, the low ionization lines come from gas photoionized by the
shock radiation. The densities we derive from the line ratios are
consistent with the large compression expected in a radiative shock.

The emission at this epoch is dominated by Spot 1. In a subsequent
paper we will discuss the evolution of these lines with time as more
spots emerge and the shocked gas at this epoch has had more time to
cool. 


\begin{acknowledgements}
We are grateful to the observers and the staff at Paranal for performing the observations at ESO VLT. This work has been supported by grants from
the Swedish Research Council and the Swedish National Space Board.
\end{acknowledgements}


\appendix
\section{Test of flux calibration}
\begin{table*}
\caption{Comparison of fluxes from HST/STIS and UVES data sets.}%
\label{t:Rep1}
\begin{tabular}{lccccc}
\hline\hline
Lines & Grism & \multicolumn{2}{c}{Flux (ergs s$^{-1}$ cm$^{-2}$)} & Ratio  & Date of HST/STIS observations \\
   &   & STIS & UVES & STIS/UVES &  \\
\hline
{H${\gamma}~\lambda$4340}          & G430L & $1.06\times10^{-14}$ &$5.74\times10^{-15}$ &1.85 &2000-11-03 \\
{[O~I]~$\lambda$6300}              & G750L & $1.20\times10^{-14}$ &$5.80\times10^{-15}$ &2.05 &2000-11-03 \\
{[S~II]~$\lambda\lambda$4068,4076} & G430L & $2.35\times10^{-14}$ &$9.83\times10^{-15}$ &2.39 &2002-10-29 \\
{H${\beta}~\lambda$4861}           & G430L & $7.19\times10^{-14}$ &$2.73\times10^{-14}$ &2.63 &2002-10-29 \\
{[N~II]~$\lambda$5755}             & G750L & $2.44\times10^{-14}$ &$1.01\times10^{-14}$ &2.41 &2002-10-29 \\
{He I~$\lambda$7065}               & G750L & $7.59\times10^{-15}$ &$3.56\times10^{-15}$ &2.13 &2002-10-29 \\
\hline
\end{tabular} \\
\end{table*}
To test our flux calibration of the UVES spectra, we downloaded archival 
HST/STIS data for two epochs: 2000 November 3 and 2002 October 29. The epochs
are close to our UVES data from 2000 December $9-14$ (discussed in 
Gr\"oningsson et al. 2007 in prep.) and 2002 October $4-7$ (discussed in this paper).
The STIS observations are long-slit spectroscopy observations made with the 
first-order gratings G430L and G750L which cover the spectral 
intervals $2900-5700$~\AA\ (resolution 2.73 \AA~pixel$^{-1}$) 
and $5250-$10\,300~\AA\ (resolution 4.92 \AA~pixel$^{-1}$), respectively. 
The aperture was $52\arcsec\times2\arcsec$, and the slit was placed so that it 
covers half of the ring, with the dispersion direction in the north-south 
direction. A second set of observations were made for the other half of the 
ring. One therefore obtains separate spectra for the northern and southern 
parts of the ring of \snr. These have to be added together to get the total 
ring flux. There is a marginal spatial overlap between the two sets of 
observations, but this is small enough (i.e., a few per cent) to be unimportant
for our test.

The pipeline-reduced STIS spectra have several cosmic rays, but there are
fortunately ``clean" lines in most parts of the spectrum so that we can
test the flux calibration for the various spectral settings of UVES. The 
lines we chose in the red part of the spectrum (G750L) are 
{[N~II]~$\lambda$5755}, {[O~I]~$\lambda$6300} and {He~I~$\lambda$7065}, and 
in the blue (G430L) we selected {[S~II]~$\lambda\lambda$4068,4076}, 
H$\gamma$ and H$\beta$.

To obtain the flux of a particular line in the STIS spectrum, we first
manually cleaned the region around the line from cosmic rays. We then
integrated the spectrum in the dispersion direction for each spatial row 
that cuts across the ring. With the pixel scale of STIS, this means that we 
made spectral scans for up to 37 spatial rows (the exact number depends on the 
brightness of the ring for the line analyzed) to cover each of the full half 
ring. We then summed up the flux from all the rows in the spectrum and added 
the flux from the opposite side of the ring to get the total ring flux for
each line. 

As the spectral resolution of STIS does not allow us to
separate the narrow component from the intermediate-velocity component,
the flux estimate from STIS is a sum of both these components. The STIS data
do, however, allow us to separate out any underlying continuum. For the
UVES spectra we therefore integrated the emission in the narrow 
plus intermediate velocity component to make a direct comparison with the STIS 
flux.

The STIS fluxes for the whole ring and ratios of STIS to UVES fluxes are 
summarized in the Table~\ref{t:Rep1}. The fact that the STIS flux is higher
than the UVES flux is reasonable since the UVES slit only encapsulates 
part of the ring whereas the STIS flux is for the full ring. The ratio
of STIS/UVES flux appears to be somewhat lower for 2000 November ($\sim 2.0$) 
than for 2002 October ($\sim 2.4$). This could be due to temporal flux 
variations round the ring (e.g., new spots appearing between 2000 and 2002), 
or may be due to a systematic error in the fluxing of the UVES spectra. The 
former is not unlikely since Spot 1, encapsulated
by the UVES aperture, dominates the ring interaction emission in the 2000 
November data, whereas two years later, it is less dominant and various parts
(also those outside the UVES slit) contribute to the ring interaction 
emission. Because the STIS/UVES flux ratio does not show any obvious 
wavelength-dependent trends (see Table~\ref{t:Rep1}), a constant 
multiplicative factor can be used for the UVES spectrum to estimate the 
total flux from the full ring. As we have discussed, the factor may be 
somewhat different for different epochs. As we here concentrate on our UVES 
data from 2002 October, we have chosen the factor 2.4 for this epoch.

\begin{figure*}[t]
\begin{center}
\includegraphics[height=85mm,angle=0, clip]{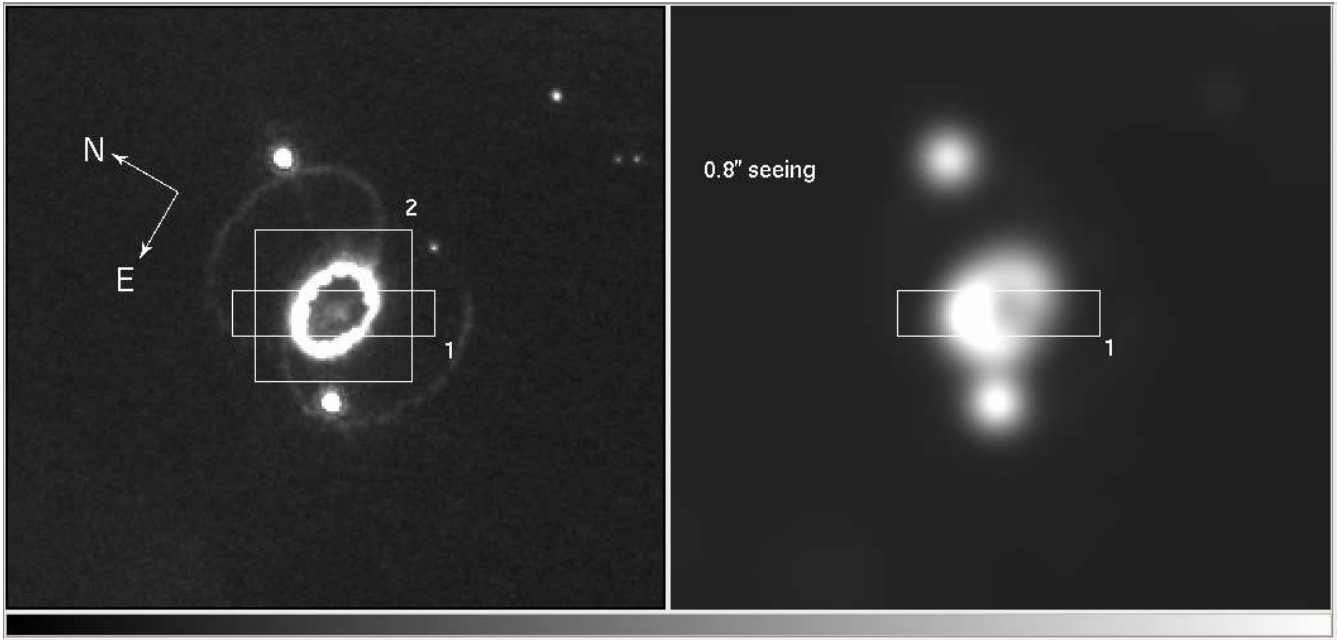}
\end{center}  
\caption{H$\alpha$ image of \snr~obtained with HST/WFPC2 in 2002 May 10 
using the narrow F656N filter centered at the rest wavelength of H$\alpha$. 
The rectangular boxes are apertures which were chosen to model UVES data 
(marked ``1'') and STIS data (marked ``2''). The image to the left is 
unconvolved. The other image is smoothed using 2D Gaussian functions and 
represent a seeing of 0$\farcs$8. The lower flux cut in both images are
roughly at the flux level of the outer rings in the image to the right.
As the inner ring is nearly two orders of magnitude brighter (cf. Fig. 1)
and the flux from Stars 2 and 3 fall below this flux cut at the position of
the UVES slit, these stars have no important influence on the flux 
calibration. Note the dominance of Spot 1 in the north-east part of the ring.
} \label{f:Ha}
\end{figure*}

To check the significance of the multiplicative factor, we downloaded 
HST/WFPC2 archival data from 2002 May 10 obtained using the F656N 
filter. The pivot wavelength for this filter is 6563.8~\AA~and the band 
width is 53.8~\AA. We convolved the image with a 2D Gaussian function to 
simulate the seeing of our UVES data. We tested three FWHM values of the 
Gaussian: 0$\farcs$6, 0$\farcs$8 and 1$\farcs$0. The Gaussian was truncated 
at 5$\arcsec$ as the inclusion of emission outside this radius did not 
change the results. Figure~\ref{f:Ha} shows the original and a blurred 
image, simulating 0$\farcs$8 seeing, with the 0$\farcs$8 UVES slit overlaid 
on top of them. The total flux from the ring was assumed to be the flux 
within the larger box in the original image. (The contributions from the 
ejecta and outer rings are low in comparison with that of the inner ring.)
We then formed a ratio of emission through the 0$\farcs$8 slit and the total
emission from the ring. The results are shown in the Table~\ref{t:Rep2}.

For the original image (i.e., no seeing), no emission from the north-west and 
south-east parts of the ring enters the 0$\farcs$8 slit. With 
increased ``seeing" emission from these parts ventures into the slit, 
whereas emission within the slit in the original image
falls outside the slit. According to Table~\ref{t:Rep2} the ``loss" of flux 
measured through the slit is slightly larger than the ``gain" since the 
ratio (total flux to UVES slit flux) increases with the degree of blurring. 
The effect is, however, rather small, with the flux within the UVES slit
decreasing from 43.7\% to 40.8\% with seeing increasing from 0$\farcs$6 
to 1$\farcs$0. This means that seeing effects have no major impact on the 
absolute fluxing of the UVES spectra, and that spectra obtained under
different seeing conditions can be coadded without introducing systematic
error in the fluxing. Also, since the ratios in Table~\ref{t:Rep2} show very 
good agreement with the ratio of 2.4 obtained from the comparison between
UVES and STIS fluxes, we are confident in the magnitude of this factor. The 
main effect of seeing is a redistribution of flux into and out from the
area covered by the slit. For a time series of spectra it therefore makes
sense to compare spectra with roughly the same average seeing to study
changes of physical properties of the ring.


\begin{table}[h]
\caption{Comparison of fluxes from HST/STIS and UVES data sets using HST/WFPC2 with F656N filter
centered at the rest wavelength of H$\alpha$. The apertures shown on the Figure~\ref{f:Ha}.}%
\label{t:Rep2}
\begin{tabular}{ll}
\hline\hline
Modeled seeing& Ratio of aperture ``2'' to ``1'' \\
\hline
unblurred  & 2.31\\
0$\farcs$6 & 2.29\\
0$\farcs$8 & 2.35\\
1$\farcs$0 & 2.45\\
\hline
\end{tabular} \\
\end{table}

\end{document}